\documentclass[]{aa}

\newcommand{\mnras}[1]{MNRAS}
\newcommand{\apj}[1]{ApJ}
\newcommand{\apjs}[1]{ApJS}
\newcommand{\apjl}[1]{ApJL}
\newcommand{\nat}[1]{Nature}
\newcommand{\aap}[1]{A\&A}
\newcommand{\aj}[1]{AJ}
\newcommand{\araa}[1]{ARA\&A}
\newcommand{\aaps}[1]{A\&ASS}
\newcommand{\apss}[1]{AP\&SS}

\usepackage{graphics}

\begin{document}

\sloppy

\thesaurus{02(12.04.1; 12.07.1; 11.08.1; 11.09.4; 11.19.2; 11.19.6)}

\title{Microlensing of multiply--imaged compact radio sources}
\subtitle{Evidence for compact halo objects in the disk galaxy 
of B1600+434}

\author{L.V.E. Koopmans\inst{1} \and A.G. de Bruyn\inst{2,1}}

\institute{Kapteyn Astronomical Institute, P.O. Box 800, NL--9700 AV
           Groningen, The Netherlands \and NFRA, P.O. Box 2, NL-7990
           AA Dwingeloo, The Netherlands}

\mail{leon@astro.rug.nl} 
\offprints{L.V.E. Koopmans}

\authorrunning{L.V.E. Koopmans \and A.G. de Bruyn} 
\titlerunning{Radio microlensing in B1600+434}
\maketitle

%%%%%%%%%%%%%%%%%%%%%%%%%%%%%%%%%%%%%%%%%%%%%%%%%%%%%%%%%%%%%%%%%%%%%%

\begin{abstract}

We present the first unambiguous case of external variability of a
radio gravitational lens, CLASS B1600+434. The {\sl Very Large Array}
(VLA) 8.5--GHz difference light curve of the lensed images, taking the
proper time-delay into account, shows the presence of external
variability with 14.6--$\sigma$ confidence.

We investigate two plausible causes of this external variability:
scattering by the ionized component of the Galactic interstellar
medium and microlensing by massive compact objects in the bulge/disk
and halo of the lens galaxy. Based on the tight relation between the
modulation-index (fractional rms variability) and variability time
scale {\sl and} the quantitative difference between the light curves
of both lensed images, we conclude that the observed short-term
variability characteristics of the lensed images are incompatible with
scintillation in our Galaxy. This conclusion is strongly supported by
multi--frequency {\sl Westerbork Synthesis Radio Telescope} (WSRT)
observations at 1.4 and 5\,GHz, which are in disagreement with
predictions based on the scintillation hypothesis.  Several arguments
against scintillation might need to be reevaluated if evidence is
found for significant scatter-broadening of lensed image B seen
through the lens galaxy. However, the frequency-dependence and
time scale of variability from image A are not affected by this and
remain strong arguments against scintillation.

On the other hand, a single superluminal jet-component in the source, 
having an apparent velocity 9$\la$$(v_{\rm app}/c)$$\la$26, a radius of
2--5\,$\mu$as and containing 5--11\% of the observed 8.5--GHz source
flux density, can reproduce the observed modulation-indices and
variability time scale at 8.5\,GHz, when it is microlensed by compact
objects in the lens galaxy. It also reproduces the
frequency-dependence of the modulation-indices, determined from the
independent WSRT 1.4 and 5--GHz observations. The difference between
the modulation-indices of the lensed images (i.e. 2.8\% and 1.6\% at
8.5\,GHz in 1998 for images A and B, respectively), if not affected by
scatter-broadening of image B by the ionized ISM of the lens galaxy,
can be explained through a different mass-function for the compact
objects in the bulge/disk and halo of the lens galaxy. Comparing the
observations with microlensing simulations, we place a tentative lower
limit of $\ga$0.5\,M$_\odot$ on the average mass of compact objects in
the halo line-of-sight. The above-mentioned set of mass-function and
source parameters is consistent, although not unique, and should only
be regarded as indicative.

The only conclusion fully consistent with the data gathered thus far
is that we have indeed detected {\sl radio microlensing}. The far
reaching consequence of this statement is that a significant fraction
of the mass in the dark--matter halo at $\sim$6\,kpc ($h$=0.65) above
the lens--galaxy disk in B1600+434 consists of massive compact
objects.


\end{abstract}

\section{Introduction}

\begin{figure*}[t!]
\resizebox{\hsize}{!}{\includegraphics{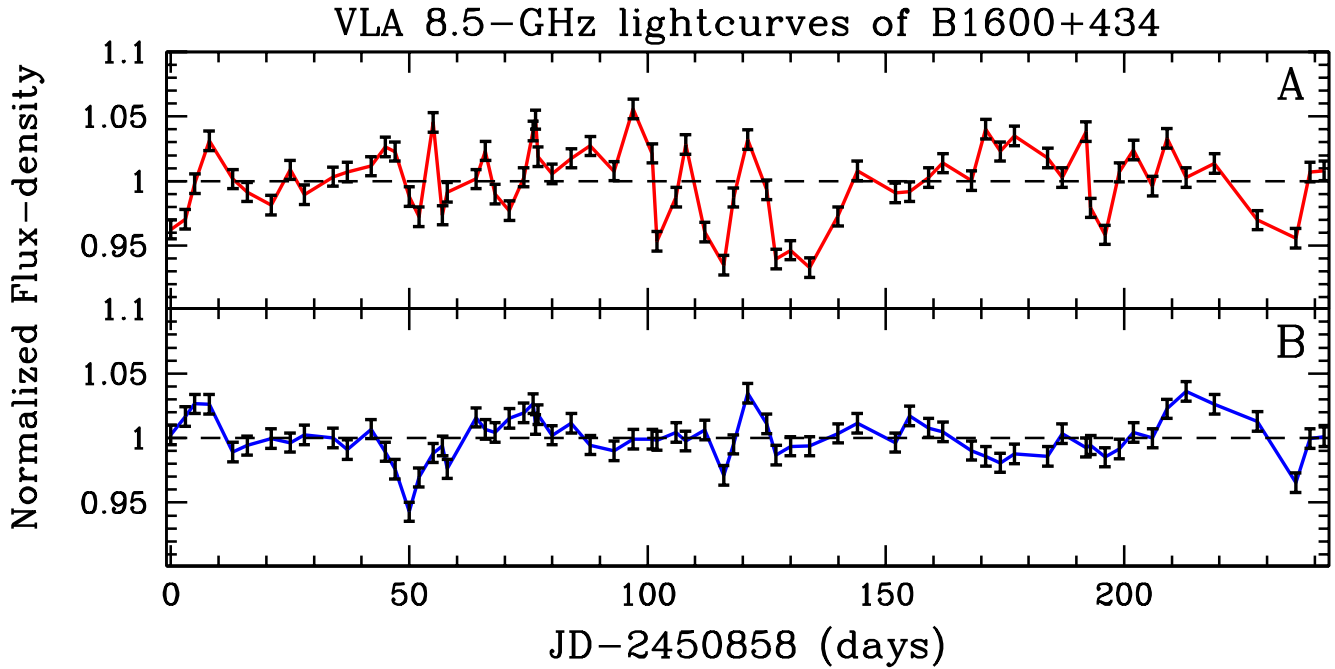}}
\hfill
\parbox[b]{\hsize}{
\caption{The normalized light curves of B1600+434 A (upper) and B
(lower), corrected for a long-term gradient (Sect.\,2). The error on
each light-curve epoch is 0.7 to 0.8\% (1--$\sigma$). Day 0
corresponds to 1998 Febr. 13}
\label{normlc}}
\end{figure*}

Gravitationally lensed compact radio sources have many astrophysical
and cosmological applications.  The foremost being the determination
of a time-delay between the individual lensed images in order to
constrain the Hubble parameter (e.g. Refsdal 1964). Considerable
progress has been made during the last few years in measuring
time-delays, both through optical and radio observations (e.g Kundi\'c
et al. 1997; Schechter et al. 1997; Lovell et al. 1998; Biggs et
al. 1999; Fassnacht et al. 1999; Koopmans et al. 2000). They also
allow a detailed study of the mass distribution of the lens galaxy and
sometimes the background source, through a large magnification by the
lensing potential.  Absorption lines in the spectrum of the background
source allow the study of the ISM in the lens galaxy and the HI
distribution along the lines of sight to the source.

Temporal changes in the brightnesses or spectra of the lensed images
also allow the study of uncorrelated external variability.  The most
important sources of external variability are scintillation at radio
wavelengths and microlensing in all wavelength bands. Differencing the
image light curves, taking the proper time delay into account, removes
intrinsic source variability and leaves only uncorrelated external
variability. These difference light curves thus provide valuable
information on the compact objects in the lens galaxy (e.g. Schmidt \&
Wambsganss 1998) and/or on the intervening ionized medium.

The study of the ionized component of the Galactic interstellar medium
(ISM) through scattering of radio emission from pulsars has had a long
tradition (e.g. Rickett 1977, 1990). Scattering by the ionized ISM can
explain long-term variability at meter wavelengths (e.g. Condon et
al. 1979), as well as large-amplitude variability in very compact
extra-galactic radio sources (e.g. Rickett, Coles \& Bourgois
1984). Low-amplitude variability at shorter wavelengths (about
10\,cm), called `flickering', has been observed by Heeschen (1982,
1984) and is probably associated with refractive interstellar
scattering of an extended source (e.g. Rickett et al. 1984).  Strong
intra-day variability of very compact radio sources might result from
refractive interstellar scattering as well (e.g. Wagner \& Witzel
1995).  A power-law model of the plasma-density power spectrum
(e.g. Rickett 1977, 1990), combined with some distribution of this
plasma in our galaxy (e.g. Taylor \& Cordes 1993 [TC93]) is able to
explain most of the observed dispersion measures and variability in
pulsars at low frequencies, as well as the variability of
extra-galactic radio sources at both low and high
frequencies. However, especially for compact flat-spectrum radio
sources it remains exceedingly difficult to separate intrinsic
variability from scintillation by the Galactic ionized ISM.

Gravitationally lensed (i.e. multiply-imaged) flat-spectrum compact
radio sources could offer a solution to this problem. As mentioned
previously, these systems provide two or more lines-of-sight through
the Galactic ionized ISM. For typical image separations of a few
arcseconds, one is looking through parts of the Galactic ionized ISM
separated by a few hundred AU. One can expect the scattering of radio
waves to be independent for the different lines-of-sight. Differencing
the image light curves, after a correction for the appropriate time
delay and flux-density ratio, produces a difference light curve that
only shows uncorrelated external variability. This difference light
curve can be studied to obtain information on the Galactic ionized ISM
independent from intrinsic source variability.

However, uncorrelated external variability of the lensed images might
also originate from microlensing in the lens galaxy (e.g. Chang \&
Refsdal 1979). This offers the additional opportunity to study the
properties of compact objects in the lens galaxy, if microlensing
variability dominates or can be separated from scintillation.  Optical
microlensing in the lens galaxy of Q2237+0305 has unambiguously been
shown (e.g. Irwin et al. 1989; Corrigan et al.  1991; Ostensen et
al. 1996; Lewis et al. 1998; Wo\'zniak et al. 2000). In the radio,
several suggestions of microlensing variability have been made
(e.g. Stickel et al. 1988; Quirrenbach et al. 1989; Schramm et
al. 1993; Romero et al. 1995; Chu et al. 1996; Wagner et al. 1996;
Lewis \& Williams 1997; Takalo et al. 1998; Quirrenbach et al. 1998;
Kraus et al. 1999; Watson et al. 1999). In none of these cases,
however, has one really been able to convincingly distinguish between
intrinsic and external variability. Claims of external variability in
singly-imaged radio sources through microlensing should therefore be
regarded with some caution.

In this paper, we report the first unambiguous case of external
variability of a radio gravitational lens, CLASS B1600+434 (Jackson et
al. 1995; Jaunsen \& Hjorth 1997; Koopmans, de Bruyn \& Jackson 1998
[KBJ98]; Koopmans et al. 2000 [KBXF00]). The system consists of two
compact flat-spectrum radio images, separated by 1.4\,arcsec.  The
background source, at a redshift of $z$=1.59, is lensed by an edge-on
disk galaxy at a redshift of $z$=0.41 (Fassnacht \& Cohen 1998). 
A time delay of $47^{+12}_{-9}$\,days (95\% statistical confidence) 
was recently found (KBXF00).

What is furthermore of interest is that this system offers two
distinct lines-of-sight through the lens galaxy. Image~A only passes
mainly through the dark-matter halo around the edge--on lens galaxy,
whereas image~B passes predominantly through its disk and bulge
(Koopmans et al. 1998; Maller et al. 2000; CASTLE Survey, Munoz et
al. 1999).  This makes image~A especially sensitive to microlensing by
massive compact objects in the halo and image~B to microlensing by
stars in the disk and bulge. This might even offer an opportunity to
study compact objects in the dark--matter halo around the lens galaxy
of B1600+434.

\begin{figure*}[t!]
\resizebox{\hsize}{!}{\includegraphics{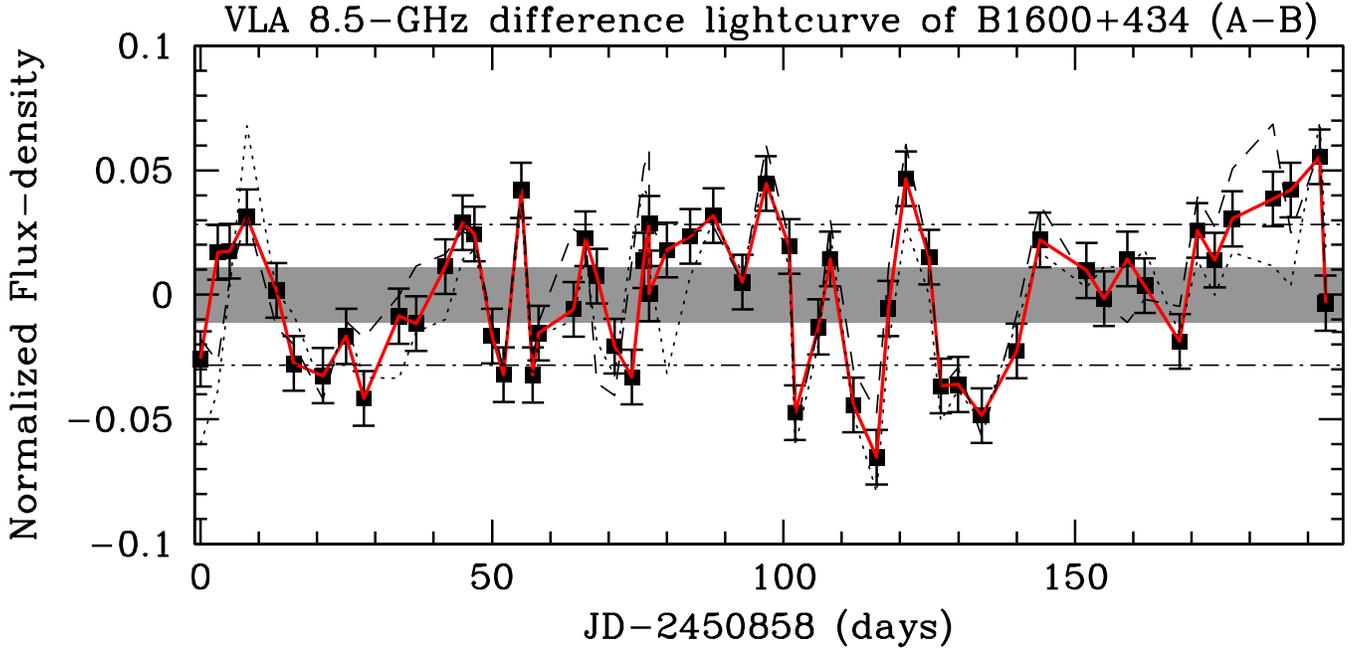}}
\hfill
\parbox[b]{\hsize}{
\caption{The normalized difference light curve between the two lensed
images, corrected for both the time-delay and flux density ratio
(Sect.\,2.1).  The shaded region indicates the expected 1--$\sigma$
(1.1\%) region if all variability were due to measurement errors. The
dash--doted lines indicate the observed modulation-index of 2.8\%. The
dotted and dashed curves indicate the normalized difference curves for
a time delay of 41 and 52\,days, respectively. Obviously most
variations can not be explained by any reasonable error in the time
delay.  There seems to be no evidence of any change in the typical
time scale of variability over 6 months. Day 0 corresponds to 1998
Febr. 13.}
\label{normdiff}}
\end{figure*}

The outline of the paper is as follows. In Section 2, we present the
VLA 8.5--GHz data from KBXF00 in a different way, unambiguously
showing the presence of external variability. We also present
additional WSRT 1.4 and 5--GHz monitoring data of B1600+434. In
Section 3, we investigate whether Galactic scintillation can explain
the fractional rms variabilities (modulation-indices) and time scales
of the short-term variability seen in the VLA 8.5--GHz light
curves. Similarly, in Sections 4 and 5 the possibility of microlensing
by compact objects in the lens galaxy is studied. In Section 6, we
present microlensing simulations of a more complex jet structure and
compare the results to B1600+434. In Section 7, we discuss a critical
test (i.e. the frequency-dependence of the modulation-index) to
discriminate between scintillation and microlensing {\sl and} compare
predictions from the VLA 8.5--GHz light curves with the independent
multi-frequency WSRT data. In Section 8 our results and conclusions
are summarized.

\section{Short-term variability in B1600+434--A \& B}

B1600+434 is a compact ($\la$1 mas at 8.5\,GHz; KBXF00) radio source,
which has varied strongly at 8.5\,GHz since its discovery in 1994
(Jackson et al. 1995). Its flux density decreased from 58 (48) mJy in
March 1994 to only 29 (24) mJy in August 1995 for image A (B)
(KBJ98). From February to October 1998, another decrease from 27 (24)
to 23 (19) mJy was found (KBXF00). In June 1999, the flux densities
appear to have stabilized to 23 (17) mJy. Strong variability was also
observed at 5\,GHz, where a total flux density was measured of 34--37\,mJy 
in 1987 (GB87; Becker, White and Edwards 1991). Observations in
March 1995 gave 45 (37) mJy for image A (B) (KBJ98), whereas in June
1999 this had reduced to only 23 (18) mJy. At 1.4\,GHz, the integrated 
WSRT flux density of B1600+434 has decreased from 60--65\,mJy in
April--July 1996 (KBJ98) to about 50\,mJy in June 1999.

Because of this strong variability, B1600+434 was observed from
February to October 1998 with the VLA at 8.5\,GHz, in order to measure
a time delay between the two lensed images. In KBXF00, the VLA
8.5--GHz light curves of the two lensed images were presented, showing
short-term variability, as well as a long-term decrease in the flux
density of both lensed images, which we assume to be intrinsic source
variability, similar to that seen in for example 0957+561 (e.g.
Haarsma et al. 1997).

In Fig.\,\ref{normlc}, the normalized VLA 8.5--GHz light curves of both
lensed images are shown. The curves were created by dividing the light
curves by linear fits (KBXF00). The resulting curves show the
fractional variability on short time scales (i.e. shorter than the
observing period of about 8 months) with respect to the running mean.
The 1--$\sigma$ error for each point on the light--curve is 0.7 to
0.8\% (KBXF00). We omitted the six strongest outliers from both plots,
which show clear systematic problems with the data or calibration
(KBXF00). The resulting normalized light curves have modulation
indices (i.e. fractional rms variabilities) of 2.8\% and 1.6\% for
images A and B, respectively. We use these values throughout this
paper. Lensed images A and B have modulation-indices significantly
larger than expected on the basis of the measurement errors only,
indicating the clear presence of short-term variability, either
intrinsic or external. The structure functions (e.g. Simonetti,
Cordes, \& Heeschen 1985) of the normalized light curves are shown in
Fig.\,\ref{strucfunc} (Sect.\,3.2).

\subsection{Intrinsic or external variability?}

To show that most of the short-term variability is of external origin,
the linearly-interpolated light curve of image~B was subtracted from
the light curve of image~A, taking a flux-density ratio of 1.212 and a
time delay of 47 days into account (KBXF00). The time delay was
determined with the minimum-dispersion method from Pelt et
al. (1996). Hence, the modulation-index of the normalized difference
light curve, shown in Fig.\,\ref{normdiff}, is by definition a lower
limit.  This is illustrated by the dotted and dashed lines, which show
the normalized difference light curves for a time delay of 41\,d and 52\,d
(i.e. the 68\% confidence region), respectively. Both curves have
larger modulation-indeces, as should be expected.

The normalized difference light curve has a modulation-index of 2.8\%.
This is significantly larger than the modulation-index of 1.0--1.1\%
(i.e. the shaded region in Fig.\,\ref{normdiff}), which one would expect
on the basis of the measurement errors only. Most of the short-term
variability present in the light--curves must therefore be of external
origin.  A $\chi^2$--value of 377 was determined from the 58 points
composing the difference light curve. This is inconsistent with a flat
difference light curve at the 14.6--$\sigma$ confidence level. 

If the short--term variability of the two lensed images is uncorrelated,
the expected modulation-index in the normalized difference curve
should be $\sqrt{(2.8^2 + 1.6^2)}$$\approx$3.2\%, which is slightly
larger than the observed value of 2.8\%. It remains hard to assess whether
individual features in the light curves might be of intrinsic
origin. In this paper we will therefore assume that {\sl all}
short--term variability in both lensed images is of external origin.

\subsection{WSRT 1.4 \& 5--GHz monitoring}

\begin{figure*}[t!]
\resizebox{\hsize}{!}{\includegraphics{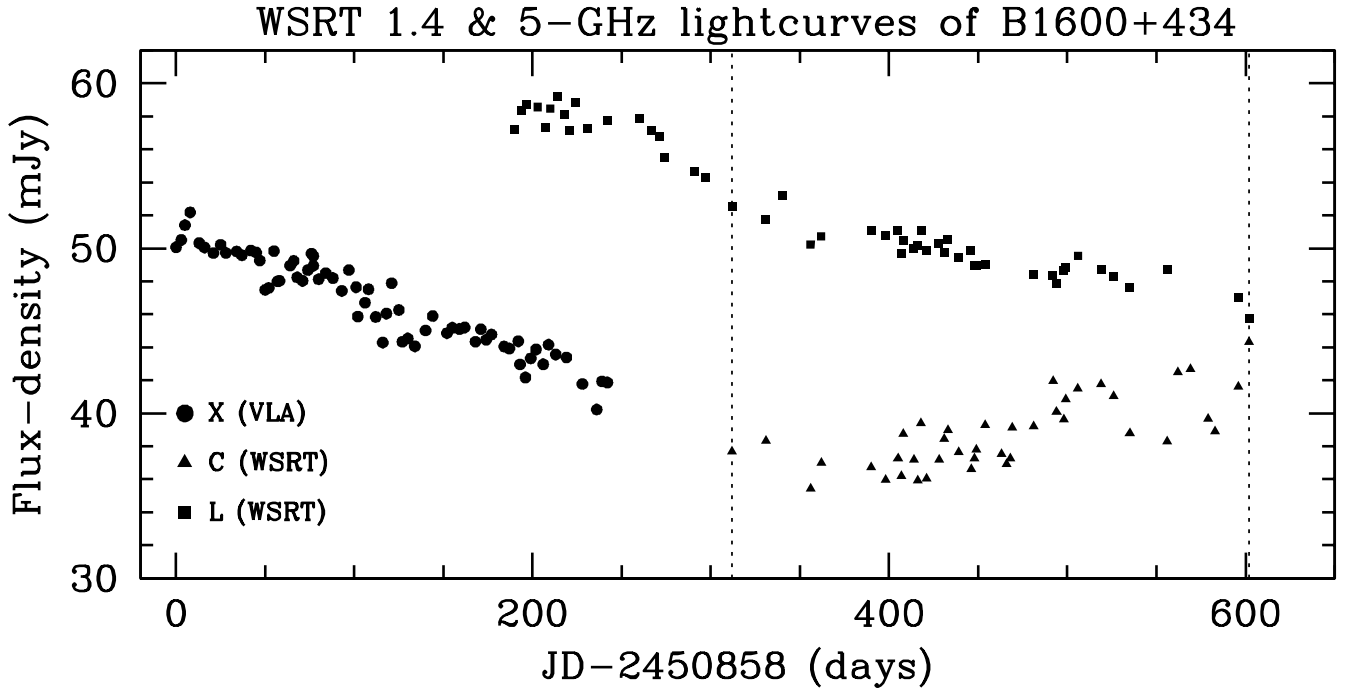}}
\hfill
\parbox[b]{\hsize}{
\caption{The integrated flux-density light curves of B1600+434 (sum of
lensed images A and B) in X--band (VLA 8.5\,GHZ), C--band (WSRT 5\,GHz)
and L--band (WSRT 1.4\,GHz). The two vertical dashed lines indicate
the time-span, where the WSRT 1.4 and 5--GHz observations
overlap. Day 0 corresponds to 1998 Febr. 13}
\label{WSRTVLA}}
\end{figure*}

Before proceeding with the analysis of the VLA 8.5--GHz light curves,
we first present multi-frequency WSRT total-flux-density data of
B1600+434 obtained in 1998/9 at 1.4 and 5\,GHz. This data will play an
essential role in distinguishing between the different physical
mechanisms causing the external variability observed in the VLA
8.5--GHz lensed-image light curves, as we will see in Sect.\,7.

Starting in August 1998 the WSRT was outfitted with a series of new
multi-frequency front-ends (MFFE's) that can operate at frequencies
from 0.3 to 8.5\,GHz. When the WSRT observations of B1600+434 were
begun in August 1998 only 6 front-ends were available at 1.4\,GHz. The
available number of telescopes with MFFE's increased at a rate of
about 1 per month until the full array was outfitted in February
1999. Towards the end of 1998 the monitoring was extended to include 5\,GHz. 
In the spring of 1999 we also included 8.5--GHz observations.
However, the analysis of the 8.5--GHz data is still encountering some
problems and we therefore do not report on the results of the 8.5--GHz
observations here.

Each run consisted of two sets of observations (at up to 3
frequencies) on B1600+434 and the nearby reference source B1558+439.
The latter is a strong steep spectrum double radio source (0.8--arcsec
size) about 40\,arcmin north-east of B1600+434.  In each run we also
observed two primary calibration sources (3C286, 3C343 and CTD93).
Although we changed the details of the observing sequence during the
year the basic structure did not change.
 
The resolution of the WSRT is about 12$\times$18\,arcsec (1.4\,GHz) and
3.5$\times$5 arcsec (5\,GHz).  B1600+434 is therefore always unresolved at
1.4\,GHz.  At 5\,GHz the source, however, shows slight hour-angle
dependent resolution effects [the WSRT is an east-west synthesis
array, hence the instantaneous synthesized response is a fan beam
rotating clockwise on the sky].  Because the observations were
scheduled at random hour angles the resolution effect is therefore
variable from session to session.  We minimized the magnitude of this
effect by determining the flux density using only baselines up to 
1300~meter; the residual effect on the flux density is below the
thermal noise error.

The amplitude and flux calibration was performed in {\sf NEWSTAR} using
standard procedures starting with a selfcalibration on the primary
reference source.  The complex telescope gains were then transferred
to the target and reference source. The flux densities of B1600+434
and B1558+439 were then determined using a uv-plane fitting algorithm
in the program {\sf NMODEL}.  Selfcal phase solutions on the reference
source B1558+439 usually showed only very slight decorrelation effects
due to slow instrumental/atmospheric phase-drifts. Because these would
be very similar for B1600+434 and the reference source we decided not
to apply a phase selfcal solution.

The flux density of the reference source B1558+439 was found to show a
scatter of about 1--2\% around 700\,mJy at 1.4\,GHz and about 1.0\%
around 204\,mJy at 5\,GHz. This scatter is still larger than the noise
error on the flux density.  We believe these can be attributed to
small changes in atmospheric opacity and small instrumental gain
drifts (e.g. due to pointing). Normalizing the amplitudes of
B1600+434 by those of the reference source should eliminate them.  We
therefore expect that the final errors on the flux density of
B1600+434 are determined by the thermal noise only.  Still, to be on
the safe side we adopted as a final error on the flux density of
B1600+434 the quadrature sum of the thermal noise level and a 1\%
scale error. This amounts to a typical error of 1.0\% and 1.3\% at 1.4
and 5\,GHz, respectively.

\begin{figure*}[t!]
\resizebox{\hsize}{!}{\includegraphics{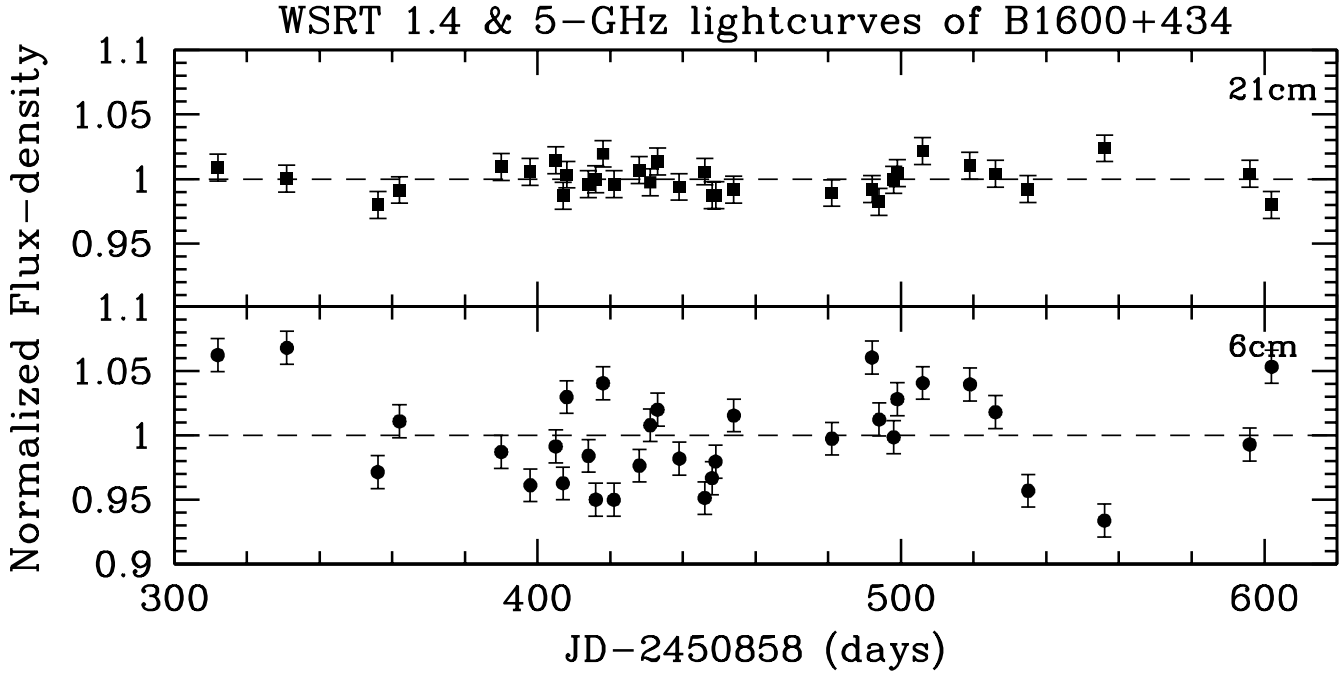}}
\hfill
\parbox[b]{\hsize}{
\caption{The normalized WSRT 1.4 and 5--GHz flux-density light curves
of B1600+434. Only those epochs of the WSRT light curves are shown
that have both a 1.4 and 5--GHz measurement (Fig.\,\ref{WSRTVLA}).  The
light curves were created similar to the VLA 8.5--GHz light curves
shown in Fig.\,\ref{normlc}. One notices a clear increase in the
modulation-index from 21 to 6\,cm by a factor $\sim$3}
\label{WSRT_norm}}
\end{figure*}

In Fig.\,\ref{WSRTVLA}, we have shown the calibrated WSRT 1.4--GHz
(L--band) and 5--GHz (C--band) light curves, together with the
total-flux-density VLA 8.5--GHz (X--band) light curve (KBXF00).  We
note the following properties: (i) All light curves are dominated by
systematic trends with decreasing intensity at 8.5 and 1.4\,GHz, but
increasing at 5\,GHz. We believe these changes to be due to intrinsic
variations. (ii) If the VLA 8.5--GHz and WSRT 5--GHz long-term
intrinsic flux-density variations are correlated, there has been a
clear trend-break in the gradient of the light curves around day 300
(Fig.\,\ref{WSRTVLA}). The WSRT 1.4--GHz light curve still shows a
similar gradient as the VLA 8.5--GHz light curves in 1998, hence there
appears to be a time-lag of at least 200\,days between long-term
intrinsic source variability at 8.5 and 1.4\,GHz. This trend-break is
supported by the first results from the 1999 VLA campaign (Koopmans et
al. in prep.). (iii) At those epochs where the light curves overlap,
the higher-frequency light curves show a larger short-term
modulation-index around the long-term linear gradient.

An important statistical property of the light curves is their
modulation-index on short time scales (i.e. time scales shorter than
the length of the light curves), as function of frequency. To
calculate the modulation-indices, we divide the light curves through a
linear fit (see Sect.\,2) in order to remove most of the presumably
intrinsic source variability. The results are normalized light curves,
similar to the normalized VLA 8.5--GHz light curves shown in
Fig.\,\ref{normlc}. We only calculate the 1.4 and 5--GHz modulation
indices ($m_{\rm part}$) for those epochs, where we have both a 1.4
and 5--GHz WSRT flux-density measurements. The normalized light curves
of these epochs are shown in Fig.\,\ref{WSRT_norm}. The resulting
modulation-indices are listed in Table~1. Because the modulation-index
at 1.4\,GHz is very close to the estimated flux-density error, we
regard it as an upper limit.

The total flux-density modulation-indices in Table~1 are a
combination of the individual modulation-indices of images~A and B. If
we assume that the ratio of modulation-indices are equal to those at
8.5\,GHz (i.e. $(m_{\rm A}/m_{\rm B})_{\lambda}$$\approx$2.8/1.6=1.75)
for each wavelength, we only make a slight error ($\la$15\%) compared
to the assumption that image~B does not vary at all on short time
scales. Because we will {\sl only} use modulation-index ratios (see
Sect.\,7), which are independent from these assumptions in first
order, this is of negligible importance.

\begin{table}
\begin{center}
\begin{tabular}{cccc}
\hline
$\nu$ (GHz) & $m_{\rm part}$ (\%) & Epoch (days) & Instr. \\
\hline
1.4 &  $\le$1.2 & 312--602 & WSRT  \\
5.0 &       3.7 & 312--602 & WSRT  \\
\hline
\end{tabular}
\end{center}
\caption{The short-term total flux-density modulation-indices of 
B1600+434, as function of frequency}
\end{table}

\subsection{Possible origins of the external variability}

What can be the origin of the external variability seen in the VLA
8.5--GHZ light curves, why does the modulation-index differ between the
two image light curves and what causes the short-term variability in
the WSRT 1.4 and 5--GHz light curves?  Below we have listed different
physical mechanisms which can introduce external variability in the 
flux density of compact radio sources:
\begin{enumerate}
	\item{Scintillation} 
	\begin{itemize} 
		\item{Weak scattering (Sect.\,3.1)}
		\item{Refractive scattering (Sect.\,3.2)} 
		\item{Diffractive scattering (Sect.\,3.3) } 
	\end{itemize}
	\item{Microlensing (Sect.\,4--5)}
\end{enumerate}
In the next three sections, we will investigate in detail whether one or
more of these can explain the external variability seen in the VLA
8.5--GHz light curves of B1600+434--A~\&~B. In Sect.\,7, we will
combine the conclusions from Sections 3--5 with the results from the
WSRT 1.4 and 5--GHz observations to further constrain the
scintillation and microlensing hypotheses.

%%%%%%%%%%%%%%%%%%%%%%%%%%%%%%%%%%%%%%%%%%%%%%%%%%%%%%%%%%%%%%%%%%%%%%
%	Section on scintillation interpretation of variability
%%%%%%%%%%%%%%%%%%%%%%%%%%%%%%%%%%%%%%%%%%%%%%%%%%%%%%%%%%%%%%%%%%%%%%

\section{Scintillation}

Even though the difference in modulation-index between images A and B
seems to require a considerable change in the properties of the
Galactic ionized ISM over an angular scale of 1.4\,arcsec (Sect.\,3.2),
we still proceed to investigate whether the short-term variability,
superposed on the gradual and presumably intrinsic long-term decrease
of the flux density of the lensed images, can be the result of
scintillation. We will follow the prescription of Narayan (1992) and
its numerical implementation by Walker (1998; [W98]) for the Galactic
ionized ISM model from TC93.  This assumes that the inhomogeneities of
the ionized ISM can be described by a Kolmogorov power-law spectrum
(e.g. Rickett 1977; Rickett 1990) and that the ionized ISM model from
TC93 is approximately valid. Support for the approximate validity of
the TC93 model in the direction to B1600+434 is given by the
dispersion and scattering measures of nearby pulsars, showing no
apparent deviations from this model. Also, no evidence is found in the
low-frequency (327--MHz) WENSS catalogue (e.g. Rengelink et al. 1997)
for diffuse HII emission or SN remnants that could introduce
small-scale perturbations in the Galactic ionized ISM.

Depending on the line-of-sight through the galaxy and the observing
frequency, the scattering strength of radio waves, expressed in the
scattering strength $\xi$=$r_{\rm F}/r_{\rm diff}$, can be strong
($\xi$$>$1) or weak ($\xi$$<$1), where $r_{\rm F}$ is the Fresnel
scale and $r_{\rm diff}$ is the diffractive scale (e.g. Narayan 1992).
The transition between these two regimes occurs near a transition
frequency ($\nu_0$). For B1600+434 at a Galactic latitude of
$b$=+48.6$^\circ$, we find $\nu_0$=4.2\,GHz (W98).  At the observing
frequency of 8.5\,GHz, scattering should therefore be in the weak
regime ($\xi$$\sim$0.3). The transverse velocity of the ISM with
respect to the line-of-sight to B1600+434 is determined by projecting
the velocity vector of the earth's motion on the sky as function of
time. We find a transverse velocity ($v$) between 20 and
40\,km\,s$^{-1}$.  We will therefore adopt an average value of
$v$=30\,km\,s$^{-1}$ throughout this paper. For lack of better
knowledge, we assume any intrinsic transverse motion of the scattering
medium to be zero.

\subsection{Weak scattering}

The modulation-index of a point source in the weak scattering regime
is (Narayan 1992)
\begin{equation}
        m_{\rm p}=\left(\frac{\nu_0}{\nu}\right)^{17/12}.
\end{equation}
In the simplest case that B1600+434 is a point source smaller than the
Fresnel scale of 3.9\,$\mu$as (W98), we would expect a modulation-index
around 35\%.  This is significantly larger than the observed
modulation-indices of 2.8\% for image~A and 1.6\% for image~B. Hence,
the total angular size of the images must be larger than the Fresnel
scale. This does, however, not exclude that part of the source might
still be compact.

The variability time scale for a point source is given by (Narayan 1992)
\begin{equation}
	t_{\rm p}= \frac{3.3}{v_{30}}\times
	\sqrt{\frac{\nu_0}{\nu}}\mbox{~~~h,}
\end{equation}
where $v_{30}$ is the transverse velocity of the scintillation pattern
with respect to the line-of-sight to the source in units of 30\,km\,s$^{-1}$.
Inserting the values for the observing frequency ($\nu$), the
transition frequency ($\nu_0$) and $v_{30}$=1.0, a time scale around
2\,h is found, which is much smaller than the apparent variability
time scales of days to weeks seen in a significant fraction of
the light curves (Figs \ref{normlc}--\ref{normdiff}).

If 5 to 10\% of the flux density of the source is contained in a
compact region ($\la$$\theta_{\rm F}$), the rms fluctuations decreases
to the observed modulation-index of 2--3\% for B1600+434--A and B. The
variability time scale would still remain $\sim$2\,h. We have observed
B1600+434 with the WSRT at 5\,GHz during several 12\,h periods and
find no evidence for short-term variability $\ga$2\% over a $\la$12\,h
time scale (Koopmans et al. in prep.). This excludes the posibility
that the longer time-scale variations are purely the result of
undersampled light curves. Hence, a simple compact source structure,
embedded in a more extended non-scintillating region of emission, can
not explain the observed variability. A more extended source
($\gg$$\theta_{\rm F}$) is therefore required, if we want to explain
the observed modulation-index in terms of scintillation.

In case the source is extended, with a size $\theta_{\rm s}$$>$$\theta_{\rm
F}$, both the modulation-index and variability time-scale change. The
modulation-index decreases as follows (Narayan 1992)
\begin{equation}
        m=m_{\rm p}\times \left(\frac{\theta_{\rm F}}{\theta_{\rm
        s}}\right)^{7/6},
\end{equation}
whereas the time scale of variability increases as
\begin{equation}
        t= t_{\rm p}\times \left( \frac{\theta_{\rm s}} {\theta_{\rm
        F}}\right).
\end{equation}
Combining these two equations, using the transition and observing
frequencies for B1600+434, gives the relation
\begin{equation}
	m=\left(\frac{0.90^{\rm h}}{v_{30}\cdot t}\right)^{7/6},
\end{equation}
for $t \ga t_{\rm p}\approx 2.25^{\rm h}/v_{30}$.

If the lensed source has an angular radius of about 40\,$\mu$as, the
modulation-index reduces to 2--3\%, as observed. The variability time scale
should then be around 1\,day, still significantly smaller than the
observed modulation timescale in a major part of the light curves.
Although part of the very short-term ($\la$ few days) variability
could be due to scintillation, many of the long-term variations seen
in Figs \ref{normlc}--\ref{normdiff} certainly can not be explained
this way.

The relatively nearby extragalactic radio source J1819+387 has a
modulation-index $m$$\approx$0.5 and a variability time scale
$t$$\approx$0.5--1\,hour (Dennett--Thorpe \& de Bruyn 2000). Using the
relation between $m$ and $t$ (e.g. Eq.\,(5)), this translates to a
variability time scale less than a day for a modulation-index of a few
percent, in agreement with expectations from the Galactic ionized ISM
model from TC93. Similarly, variations with a time scale of more than
a week (Figs \ref{normlc}--\ref{normdiff}) require a source size of
0.3\,mas, reducing the modulation-index to 0.2\%, which is well below
the noise level in the VLA 8.5--GHz light curves.

\begin{figure}[t!]
\center 
\resizebox{\hsize}{!}{\includegraphics{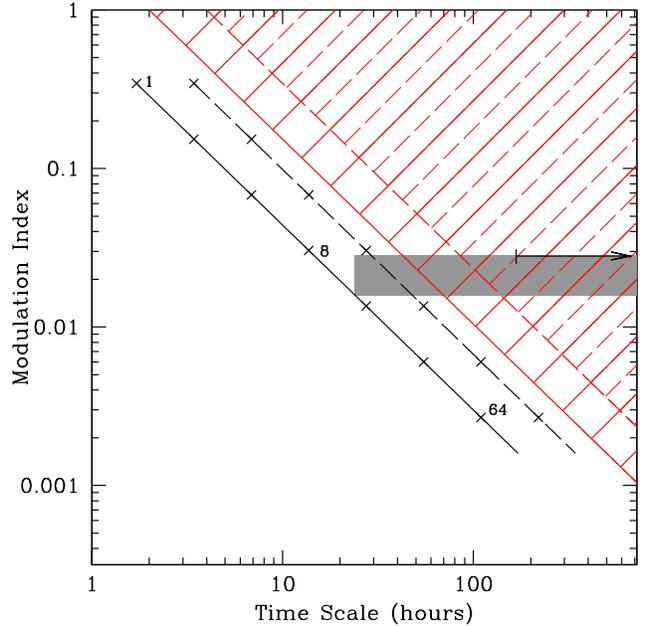}}
\vfill
\parbox[b]{\hsize}{
\caption{The modulation-index as function of the variability time
scale for the weak scattering regime (Eq.\,(5)), using a transverse
velocity $v$=40\,km~s$^{-1}$ (solid line) or 20\,km~s$^{-1}$ (dashed
line). The short dashes perpendicular to the lines indicate different
source sizes (i.e. 1, 2, 4, ..., 64 in units of the first Fresnel
zone, i.e. 3.9 $\mu$as). The skewed dashed regions in the upper-right
corner indicate the strong scattering regime ($\xi>1$) for the two
transverse velocities, using an equivalent phase-screen distance of
0.5\,kpc. The gray region indicates a variability time scale of 1--30
days and a modulation-index range of 1.6--2.8\%, as observed for the
lensed images. The arrow indicates the observed 2.8\% modulation-index
and time scale $\ge$1 week as seen in the difference light curve
(Fig.\,\ref{normdiff}). Only for an equivalent phase screen distance
$\ge$10\,kpc (with $v$=30\,km~s$^{-1}$) or a transverse velocity
$\le$few~km\,s$^{-1}$ (with $D$=0.5\,kpc) does most of the gray region
enter the weak scattering regime.}
\label{scatterreg}}
\end{figure}

In Fig.\,\ref{scatterreg}, we have summarized the weak and strong
scattering regimes, as function of the modulation-index, the
variability time scale and the source size. The variability seen in
image~A (but also image~B) is especially hard to explain by weak
scattering without either invoking unlikely high values for the
equivalent distance of the phase screen ($\ga$10\,kpc) or a
persistently low transverse velocity ($\la$few~km\,s$^{-1}$). From
this, we conclude that weak scattering has great difficulties in
accounting for the observed modulation-indices and longer variability
time scales ($\gg$1 day) seen in the VLA 8.5--GHz light curves of the
lensed images. However, it remains difficult to determine a reliable
variability time scale from the image light curves, partly because of
the relatively poor sampling (i.e.  every 3.3~days). In any case,
sigificant variability on short time scales ($\la$12\,h) is excluded
(see above). For further discussion we refer to Sect.\,3.2, where we
present the structure functions of the observed variations.

\subsection{Strong scattering}

In the strong scattering regime, we can not use the numerical results
derived from the TC93 model, from which one expects B1600+434 to be in
the weak-scattering regime at 8.5\,GHz. We therefore make direct use of the
relation between the scattering measure (SM), the distance to the
equivalent phase screen ($D_{\rm kpc}$), the observing frequency
($\nu_{\rm GHz}$) and the scattering strength ($\xi$) (e.g. W98)
\begin{equation}
        \xi=2.6\cdot10^{3}\times {\rm SM}^{6/10}
        D^{1/2}_{\rm kpc} \; \nu_{\rm GHz}^{-17/10}.
\end{equation}
The scattering strength and the Fresnel scale determine both the
modulation-index and variability time scale of a source, given the
source size. The Fresnel scale, given by
\begin{equation}
	\theta_{\rm F}=8/\sqrt{D_{\rm kpc} \nu_{\rm GHz}} ~~~\mu\mbox{as},
\end{equation}
specifies the angular distance from the source over which there is
about one radian phase difference between rays, due to the difference
in path length. The scattering measure (e.g. TC93) for an
extra-galactic source is defined as
\begin{equation}
	{\rm SM}=\int_0^{\infty} C_{\rm n}^2{\rm d}l, 
\end{equation}
where $C^2_{\rm n}$ is the structure constant normalizing the Kolmogorov
power-law spectrum of the ionized ISM inhomogeneities (e.g. Cordes, Weisberg
\& Boriakoff 1985). From now on, we assume that SM has units of kpc
m$^{-{20}/{3}}$ and $C^2_{\rm n}$  units of m$^{-{20}/{3}}$. The distance 
to the equivalent phase screen (e.g. W98) is defined as
\begin{equation}
        D\equiv \frac{1}{{\rm SM}}\int_{0}^{\infty} C_{\rm
	n}^2\cdot s \;{\rm d}s.
\end{equation}

Despite the fact that the difference in modulation-index of the lens
images seems to require very different properties of the Galactic
ionized ISM on a scale of 1.4\,arcsec, we will investigate the two
distinct strong scattering regimes, i.e. refractive and diffractive
(e.g. Rickett 1990; Narayan 1992), in more detail in the next two
sections.

\begin{figure*}
\resizebox{8.5cm}{!}{\includegraphics{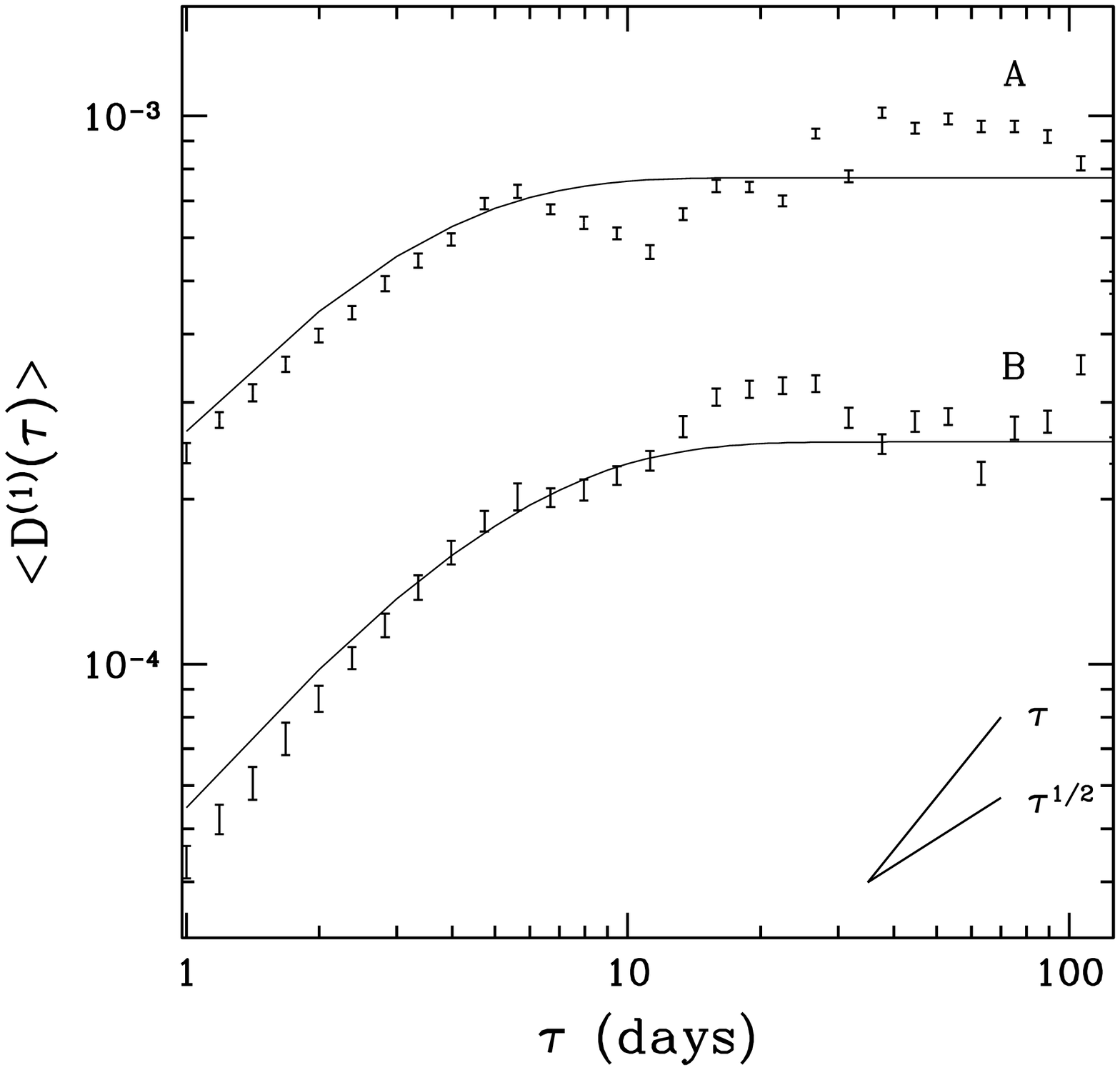}}
\resizebox{8.5cm}{!}{\includegraphics{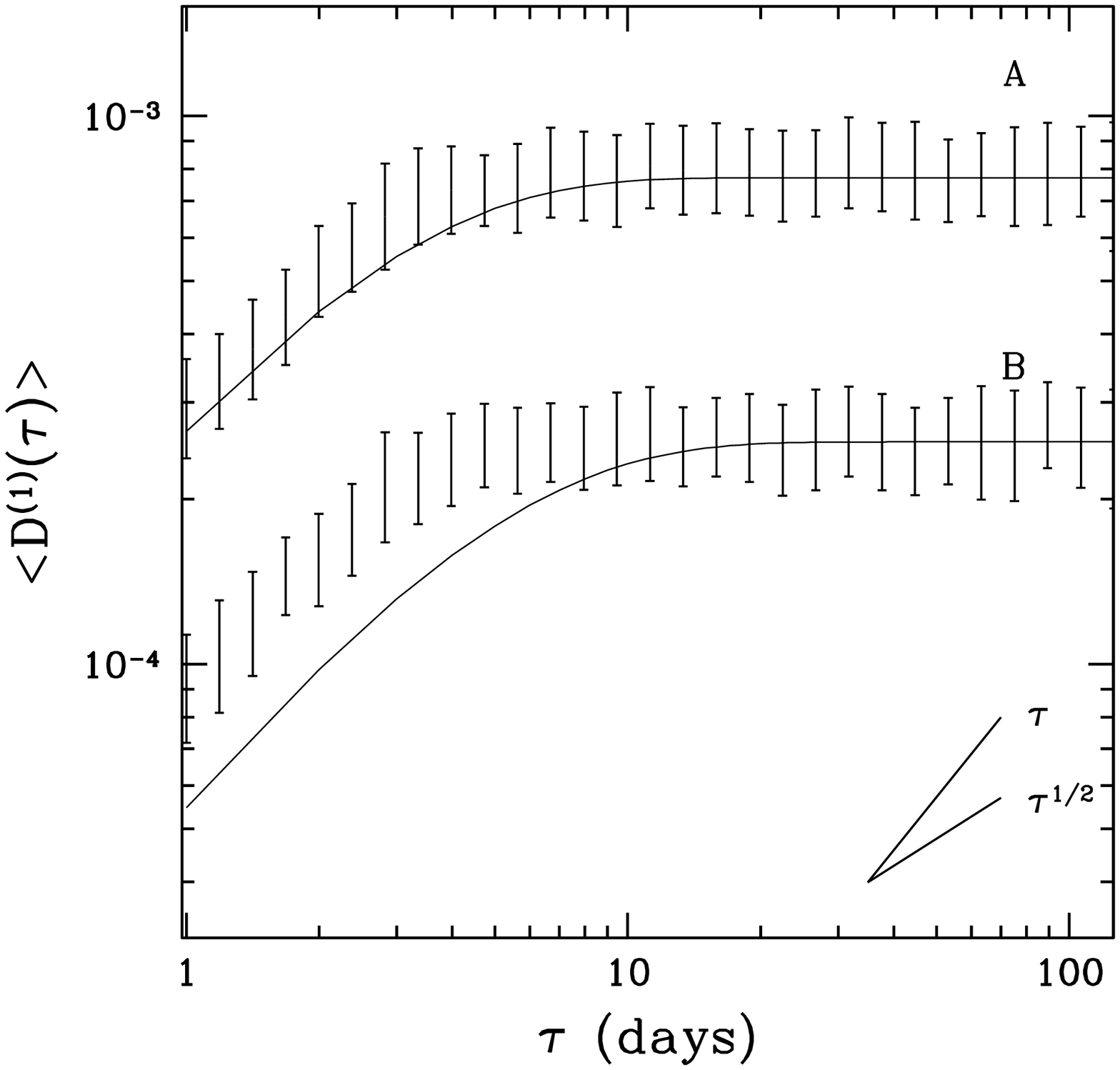}}
\vfill
\parbox[b]{\hsize}{
\caption{{\sl Left:} 
The structure functions of the normalized light curve of images~A and
B. The solid curves shows the expected structure functions, derived as
in BNR86, assuming a scale length of 0.9\,kpc for the Gaussian
distribution of the ionized scattering medium and $v$=30\,km~s$^{-1}$. 
Fits are only obtained for very different source sizes of 62 
and 108\,$\mu$as for images A and B, respectively. The
1--$\sigma$ error bars on the structure functions are derived from
Monte-Carlo simulations. {\sl Right:} The simulated structure
functions of the normalized light curve of images~A and B, replacing
the observed normalized flux densities at each epoch with a random
Gaussian distributed value with a 1--$\sigma$ scatter equal to the
observed modulation-index in the light curves, i.e. 2.8\% and 1.6\%
for images A and B, respectively}
\label{strucfunc}}
\end{figure*}

\subsubsection{Refractive scintillation}

Using Eq.\,(6) and the scaling laws from Narayan (1992), one finds for a
point source in the strong scattering regime that the modulation-index is
\begin{eqnarray}
	m_{\rm p}&=&\xi^{-{1}/{3}}\\
	&\approx& 
	7.3\cdot 10^{-2} \cdot {\rm SM}^{-{1}/{5}}\,
	D_{\rm kpc}^{-{1}/{6}}\, \nu_{\rm GHz}^{{17}/{30}},\nonumber
\end{eqnarray}
whereas the variability time scale  is
\begin{equation}
	t_{\rm p}=\frac{\theta_{\rm F} D}{v}\xi \approx 
	\frac{3.3}{v_{30}}\, D_{\rm kpc}\, {\rm SM}^{{6}/{10}}\,
	\nu_{\rm GHz}^{-{11}/{5}}\mbox{~~~yrs}.
\end{equation}

We furthermore use $D_{\rm kpc}$=0.5 (TC93), $v_{30}$=1.0 and
$\nu_{\rm GHz}$=8.5 throughout this section.  From Eqs.\,(10--11) it is
immediately obvious that for a point source in the refractive regime,
an extremely high value for SM is needed ($\ga 10^5$\,kpc~m$^{-20/3}$)
to obtain the modulation-index of images A and B. The time scale of
variability would be around 15~years. Clearly the point-source
approximation is not valid. 

For extended sources, the modulation-index and time scale of
variability scale as $(\theta_{\rm r}/\theta_{\rm s})^{{7}/{6}}$ and
$(\theta_{\rm s}/\theta_{\rm r})$, respectively, where $\theta_{\rm
s}$ is the source size and $\theta_{\rm r}$ the size of the scattering
disk (Narayan 1992).  At 8.5\,GHz, we find
\begin{equation}
	\theta_{\rm r}=\theta_{\rm F}\cdot \xi \approx 180\times {\rm
	SM}^{{6}/{10}}~~~\mu{\rm as},
\end{equation}
which is independent of the distance to the equivalent phase screen.
If we subsequently use the scaling laws, combined with Eqs.\,(10--11),
we find a relation between the time scale of variability and the
modulation-index:
\begin{equation}
	m = \xi^{{5}/{6}} \left(\frac{\theta_{\rm F} D} {v\cdot
	t}\right)^{{7}/{6}},
\end{equation}
which is valid only if $\theta_{\rm s}>\theta_{\rm r}$. Inserting
the usual numerical values for $v$, $D$ and $\nu$, we find
\begin{equation}
	m \approx 2.0\cdot {\rm SM}^{{1}/{2}}\, t_{\rm d}^{-7/6},
\end{equation}
with $t_{\rm d}$ in units of days. We find that a scattering measure
SM$\ga$$2\cdot 10^{-2}$ is needed to explain modulations with a time
scale of $\ga$1~week in image~A. From TC93 we find that
SM=$2\cdot10^{-4}$ in the direction of B1600+434, corresponding to a
time scale of one day. For deep modulations of about 1~month a
scattering measure SM$\approx$0.5 is needed for image~A.  Both values
are larger than can be expected on the basis of the
ionized ISM model from TC93.

\paragraph{\sf Differences in the scattering measure:}
A strong argument against refractive scattering is the
large difference between the modulation-indices of images~A and B.  If
this is due to a difference in the scattering measure, it requires
${\rm SM_A}/{\rm SM_B}$$\approx$$(0.028/0.016)^2$$\approx$3.1 
(Eqs.\,(13--14)), which is substantial over an angular scale of only 1.4
arcsec. The same factor is found for the weak-scattering regime.

\paragraph{\sf The structure function:}
We have also calculated the structure function (SF; Simonetti et al. 1985):
\begin{equation}
	{\rm D}^{(1)}(\tau)=\frac{1}{2}<(\delta {\rm F}(t) - \delta
	{\rm F}(t+\tau))^2>,
\end{equation}
following Blandford, Narayan \& Romani (1986 [BNR86]), who
investigated intensity fluctuations (i.e. ``flickering'') of extended
radio sources, caused by refractive scattering. ${\rm F}(t)$ is the
normalized light curve as shown in Fig.\,\ref{normlc}. BNR86 take a
slightly steeper spectrum of the phase fluctuations with a
power-spectrum slope $\beta$=4, instead of a Kolmogorov slope of
$\beta$=11/3.

Fitting the theoretical SF from BNR86 to the observed SFs\footnote{The
structure function D$^{(1)}(\tau)$, calculated from the normalized
light--curves, corresponds to the structure function D$^{(2)}(\tau)$,
calculated from the observed light curves (KBXF00), because SFs of
order $M$ remove polynomials of order $M$--1 from the time series
(e.g. Simonetti et al. 1985).}  (Fig.\,\ref{strucfunc}) gives a scale
length of $L$$\sim$0.9\,kpc, which corresponds to an equivalent phase
screen distance of 0.9$/\sqrt{\pi}\approx$0.5\,kpc (Eq.\,(9)). The
`best-fit' image sizes are 62 and 108 $\mu$as, respectively, for
images A and B. The saturation time scales (e.g. BNR86) found from
these fits are $\tau_{\rm s}^{\rm A}$=2.5~days and $\tau_{\rm s}^{\rm
B}$=4.4~days, even though there are clear variations with longer
time scales in the light curves. The presence of variabilty with longer 
time scales ($\ga$week) has been confirmed by new multi-frequency
VLA observations in 1999/2000 (Koopmans et al. in prep.).

To test the reliability of these saturation time scales, we replaced
the normalized flux densities at each epoch in Fig.\,\ref{normlc} by
Gaussian--distributed values with a 1--$\sigma$ scatter equal to the
observed modulation-index of the light curve. In Fig.\,\ref{strucfunc}
the result is shown, from which it is immediately clear that the light
curves are undersampled such that the SFs and the saturation time
scales for time lags $\tau$$\la$4 days become highly unreliable.

\paragraph{\sf Scatter-broadening:}

The difference between the modulation-indices of images A and B can be
explained by a difference in the scattering measure of the Galactic
ionized ISM towards both images, as well as by a difference in their
respective image sizes (see above). However, image B has a smaller
magnification due to the lensing potential and should therefore be
smaller than image A. Consequently, image B should show stronger
variability than image A, whereas it does not. The only viable
alternative to obtain a larger size for image B is through
scatter-broadening by the ionized ISM in the lens galaxy.

The expected scattering disk at 8.5\,GHz due to the Galactic ionized
ISM is $\sim$1\,$\mu$as and cannot account for the apparent difference
in the image size, derived from the SFs. This requires a scattering
disk of $\ga$90\,$\mu$as for image B at 8.5\,GHz, if image A is not
scatter-broadened, implying that SM$_{\rm B}$$\approx$1 in the lens
galaxy. If we take into account that scattering occurs at a frequency
of 8.5$\times(1+z_{\rm l})$$\approx$12.0\,GHz, this implies a
scattering disk of 3\,mas at 1.7\,GHz, which is nearly equal to the
very conservative upper limit of $\la$4\,mas on the image sizes found
from 1.7--GHz VLBA observations (Neal Jackson, private communication).

Recent polarization observations by Patnaik et al. (1999) gave
rotation measures RM=40\,rad~m$^{-2}$ and has RM=44\,rad~m$^{-2}$ for
images A and B, respectively. The difference of 4$\pm$4\,rad~m$^{-2}$
is rather low and certainly does not support a high scattering measure
in the disk/bulge of the lens galaxy.

Hence, although scatter-broadening cannot be excluded, to fully explain the
observed difference between the modulation-indices of images A and B in 
terms of galactic scintillation, one would require an extremely high
scattering measure in the lens galaxy.

\subsubsection{Diffractive scintillation}

For diffractive scintillation at 8.5\,GHz to be at work, one requires
both a very high scattering measure and a very small source, neither
of which seems plausible. However, to be complete we briefly discuss this
possibility.

The modulation-index is unity for a point source,
\begin{equation}
	m_{\rm p}=1,
\end{equation}
much larger than seen in both lensed images. However, for a source
larger than the scale on which there are phase changes of about 1~radian 
($\theta_{\rm d}$), the modulation-index becomes $m=(\theta_{\rm
d}/\theta_{\rm s})$, where $\theta_{\rm s}$ is the source size. We
find (e.g. W98)
\begin{eqnarray}
	\theta_{\rm d}&=&\theta_{\rm F}\;\xi^{-1}\\
	&\approx&
	3.1\cdot 10^{-3}\cdot {\rm SM}^{-{6}/{10}}
	\nu_{\rm GHz}^{{6}/{5}} D_{\rm kpc}^{-1}\mbox{~~~$\mu$as}\nonumber, 
\end{eqnarray}
and for the point--source variability time--scale
\begin{equation}
	t_{\rm d}=t_{\rm F}\cdot \xi^{-1}\approx\frac{0.25}{v_{10}}\cdot
	{\rm SM}^{-{6}/{10}}\nu_{\rm GHz}^{{6}/{5}} \mbox{~~~min},
\end{equation}
where $t_{\rm F}$=$\theta_{\rm F} D v^{-1}$=$4.0\cdot 10^4\cdot
v_{30}^{-1} \sqrt{D_{\rm kpc}/\nu_{\rm GHz}}$\,sec.  The time scale
increases by $(\theta_{\rm s}/\theta_{\rm d})$, if the source size is
larger than $\theta_{\rm d}$. Combining the equations above, we find the
relation
\begin{equation}
	\xi=\frac{t_{\rm F}}{m \cdot t},
\end{equation}
where $m$ is the observed modulation-index.  This relation is
independent of source size, as long as the source is larger than the
diffractive scale $\theta_{\rm d}$. Using this equation, we
immediately find that $\theta_{\rm d}>\theta_{\rm F}$ for the deep
modulations of $\ga$1 week, which is only true in the weak scattering
regime. Thus, diffractive scattering offers no solution either, which
comes as no surprise.

%%%%%%%%%%%%%%%%%%%%%%%%%%%%%%%%%%%%%%%%%%%%%%%%%%%%%%%%%%%%%%%%%%%%%%
%	Section on microlensing interpretation of variability
%%%%%%%%%%%%%%%%%%%%%%%%%%%%%%%%%%%%%%%%%%%%%%%%%%%%%%%%%%%%%%%%%%%%%%

\section{Radio Microlensing: Theory}

Microlensing is unlikely for bright ($\ga$1\,Jy) flat-spectrum radio
sources at low frequencies ($\sim$few~GHz), which have typical angular
sizes of $\ga$1\,mas, determined by the inverse Compton limit on their
surface brightness (e.g. Kellerman \& Pauliny--Toth 1969). This
angular size is much larger than the typical separation of a few
$\mu$as between caustics in the magnification pattern, thereby 
reducing any significant microlensing variability (see Sect.\,5).

High-frequency ($\sim$10\,GHz) sources with flux densities
less than a few tens of mJy, however, can be as small as several tens
of $\mu$as. If part of the source is moving with relativistic
velocities ($\beta_{\rm app}c$), Doppler boosting allows an even
smaller angular size. In those cases, microlensing can start to
contribute significantly to the short-term variability seen in these
sources (e.g. Gopal-Krishna \& Subramanian 1991).

This is in stark contrast to optical microlensing, where the
variability time scales are dominated by the transverse velocity
($v_{\rm trans}$) of the galaxy with respect to the line-of-sight to
the static source, as in the case of Q2237+0305 (e.g. Wyithe, Webster
\& Turner 1999). Microlensing time scales between strong caustic
crossings in the optical waveband are therefore several orders of
magnitude (i.e. $\beta_{\rm app}c/v_{\rm trans}$$\sim$$10^{3}$) larger
than in the radio waveband. This makes superluminal radio sources the
perfect probes to study microlensing by compact objects in strong
gravitational lens galaxies, using relatively short ($\la$1~year)
monitoring campaigns.

\subsection{Relativistic jet-components}

If the lensed source consists of a static core and a synchrotron
self-absorbed jet-component, which moves away from the core with a
velocity $\beta_{\rm bulk}$=$v/c$, the Doppler factor of this
jet-component is given by
\begin{equation}
  {\cal D}=[\gamma \; ( 1 - \vec{\beta}_{\rm bulk}\cdot \vec{n})]^{-1},
\end{equation}
where $\gamma$=$(1-\beta_{\rm bulk}^2)^{-1/2}$ and $\vec{n}$ is the
direction of the observer (e.g. Blandford \& K\"{o}nigl 1979).

The observed flux density of a circular-symmetric radio source with
an observed angular radius $\Delta \theta$ is
\begin{equation}
  S(\nu)=\frac{2 k T_{\rm b} \nu^2}{c^2}\cdot \pi{\Delta \theta}^2,
\end{equation}
where $\nu$ is the observing frequency and $T_{\rm b}$ the observed
brightness temperature of the source. We assume the source has a
constant surface brightness. However, due to the Doppler boosting, the
apparent brightness temperature of a relativistic jet--component moving towards
the observer can be significantly brighter than the inverse Compton
limit of about $T_{\rm b,ic}$$\approx$$5\cdot 10^{11}$K (e.g. Kellerman \&
Pauliny-Toth 1969). The true comoving surface brightness temperature
of a flat-spectrum radio source is related to the observed surface
brightness temperature ($T_{\rm b}$) through
\begin{equation}
  T_{\rm b}= 10^{12} 
	\times T^{\rm b,ic}_{12} \left(\frac{{\cal D}}{1+z}\right)\rm{~~~K},
\end{equation}
where $z$ is the redshift of the source (e.g. Blandford \& K\"{o}nigl
1979). If we substitute Eq.\,(22) into Eq.\,(21), we find that the
flux density of the relativistically moving jet-component is
\begin{equation}
  S_{\rm knot} = 0.23\left[\frac{{\cal D}\, T^{\rm
      b,ic}_{12}}{1+z}\right] \nu_{10}^2 {\left(\frac{\Delta
      \theta_{\rm knot}}{\mu{\rm as}}\right)^2}\mbox{~~~mJy},
\end{equation}
with $\nu$=10$\cdot$$\nu_{10}{\rm ~GHz}$. Inverting this equation, we find
an approximate angular radius of the jet-component
\begin{equation}
  \Delta \theta_{\rm knot} = 2.1\cdot \left[\left(\frac{S_{\rm
       knot}}{\mbox{mJy}}\right)\frac{ (1+z)}{{\cal D}\, T^{\rm
       b,ic}_{12}\nu_{10}^{2} }\right]^{1/2} \mbox{~~~}\mu\mbox{as}.
\end{equation}
Given the observed frequency, the redshift of the jet-component and
the Doppler boosting ${\cal D}$, we can subsequently set a limit on the
angular radius of the jet-component.

In the case of B1600+434, the redshift of the lensed quasar is
$z$=1.59 (Fassnacht \& Cohen 1998). The observing frequency is
$\nu_{10}$=0.85 and $S_{\rm knot}$$\approx$$S^{\rm tot}_{\rm 8.5}\cdot
f$, where $f$ is the fraction of the total average source flux density
in the relativistic jet-component, 
$S^{\rm tot}_{8.5}$$\approx$$25/\langle\mu\rangle$ mJy (KBXF00) and
$\langle\mu\rangle$ is the average magnification at the image
position.  From the singular isothermal ellipsoidal (SIE) mass model
(Kormann, Schneider \& Bartelmann 1994), we find $\langle\mu_{\rm
A}\rangle$$\approx$1.7 ($\kappa$=0.2) and $\langle\mu_{\rm
B}\rangle$$\approx$1.3 ($\kappa$=0.9). We use
$\langle\mu\rangle$$\approx$1.5 as a typical value.

After inserting all the known observables into Eq.\,(24) and adopting
$T^{\rm b,ic}_{12}$=0.5, we find an approximate relation between the
fraction of the total observed flux--density of B1600+434 contained in
the jet--component and its angular size in the source plane
\begin{equation}
        \Delta \theta_{\rm knot} \approx 20 \cdot \sqrt{\frac{f}{{\cal
        D}}}\mbox{~~~}\mbox{$\mu$as}.
\end{equation} 
This equation can be used to put constraints on the light-curve
fluctuations seen in B1600+434, and decide whether they are the result 
of microlensing of a single relativistic jet-component.

\subsection{Microlensing time scales}

The typical time scale which one would expect between relatively strong
microlensing events is the angular separation between strong caustic
crossings divided by the angular velocity of the jet-component in the source plane.

In case the source is not lensed, the apparent velocity (in units of
$c$) of the jet-component is
\begin{equation}
  \vec{\beta}_{\rm app}=\frac{\vec{n} \times (\vec{\beta}_{\rm bulk} 
	\times \vec{n})}
    {1-\vec{\beta}_{\rm bulk}\cdot\vec{n}}=
    \frac{\vec{\beta}_{\rm bulk} \sin(\psi)}{1-|\vec{\beta}_{\rm bulk}|
	\cos(\psi)},
\end{equation}
where $\psi$ is the angle between the jet and the line--of--sight and
$\beta_{\rm bulk}$ the bulk velocity of the jet--component (e.g. Blandford \&
K\"{o}nigl 1979). The apparent angular velocity (in vector notation)
of the jet--component becomes
\begin{eqnarray}
  \frac{{\rm d} \vec{\theta}_{\rm s}}{{\rm d}t}&=&
	\frac{\vec{\beta}_{\rm app} c}{(1+z_{\rm s})D_{\rm s}}\\
  &=&\frac{1.2\cdot \vec{\beta}_{\rm app}}{(1+z_{\rm s})}\left(\frac{D_{\rm s}}
	{\rm Gpc}\right)^{-1}\;\;
  \frac{\mu{\rm as}}{\rm week},\nonumber
\end{eqnarray}
where $z_{\rm s}$ and $D_{\rm s}$ are the redshift and the angular
diameter distance to the stationary core, respectively.  If the
jet-component moves with superluminal velocities ($\beta_{\rm app}\ge
1$), one expects angular velocities in the order of several tenths of
$\mu{\rm as}$ per week. Because the source is lensed by the
foreground galaxy, its observed angular velocity (in the lens plane)
becomes
\begin{equation}
   \frac{{\rm d}\vec{\theta}_{\rm d}}{{\rm d}t}= \left[{\cal A}^{-1}
	\cdot \frac{{\rm d} \vec{\theta}_{\rm s}}{{\rm d}t}\right]
	\left(\frac{D_{\rm d}}{D_{\rm s}}\right),
\end{equation}
where
\begin{equation}
	{\cal A}=\left[\begin{array}{cc} 1-\kappa-\gamma_1 & -\gamma_2
	\\ -\gamma_2 & 1-\kappa+\gamma_1\end{array}\right] 
\end{equation}
is the local transformation matrix of the source plane to the lens
plane, with $\kappa$ and $\gamma_{1,2}$ being the local convergence
and shear (e.g. Gopal-Krishna \& Subramanian 1991; Schneider et al. 1992).

We calculate the source and caustic structure in the source plane,
however. Thus, the angular velocity and source structure undergo the
inverse transformation of the angular velocity in the source plane
(Eq.\,(28)). The angular velocity that we need to use, is therefore
given by Eq.\,(27). Using the observed redshift $z$=1.59 (Fassnacht \&
Cohen 1998) of B1600+434 the angular velocity in the source plane
reduces to
\begin{equation}
      \frac{{\rm d} \vec{\theta}_{\rm s}}{{\rm d}t}= 0.34 \cdot
      \vec{\beta}_{\rm app} \;\; \frac{\mu{\rm as}}{\rm week},
\end{equation}
where we assume a flat Friedmann--Robertson--Walker universe with
$\Omega_{\rm m}$=1 and H$_0$=65 km s$^{-1}$ Mpc$^{-1}$.

\subsection{Microlensing modulation-indices}

The normalized modulation-index ($m_{\mu}$) of a superluminal jet
component, due to microlensing, is
\begin{equation}
        m_{\mu}\equiv \sqrt{\frac{\langle\mu^2\rangle}
	{\langle\mu\rangle^2} -1},
\end{equation} 
where $\mu$=$\mu(t)$ is the microlensing light curve of the jet
component.  We determine $m_\mu$ as function of the angular size of
the source by averaging over randomly-oriented simulated microlensing
light curves and find that $m_\mu$ can be well approximated by the
analytical function
\begin{equation}
        m_\mu(\Delta \theta_{\rm knot})=
	\frac{C_\mu}{1+\left(\frac{\Delta \theta_{\rm knot}}
	{\Delta\theta_{\rm b}}\right)}
\end{equation}
for $\Delta \theta_{\rm knot}$$\la$20$\mu$as, where $\Delta\theta_{\rm b}$
is the turnover size of the source after which the modulation-index
$m_\mu$ decreases linearly with source size and $C_\mu$ is the
asymptotic modulation-index for a source with $\Delta \theta_{\rm
knot}$$\rightarrow$0.  We fit this function to the numerical results
to obtain both $C_\mu$ and $\Delta \theta_{\rm b}$.

We subsequently combine Eqs.\,(25) and (32) with the fact that the observed
modulation-index in the lensed images is $m^{\rm obs}_\mu=f\cdot m_\mu$ and
find
\begin{equation}
        m^{\rm obs}_\mu(f,{\cal D}) \approx \frac{C_\mu \cdot
        f}{1+20\;\Delta\theta_{\rm b}^{-1} \; \; \sqrt{f/{\cal D}}},
\end{equation}
where $\Delta\theta_{\rm b}$ is in units of $\mu$as.

Many jets consist of more than a single jet-component.  If we assume
that (i) the jet consists of $N$ similar jet-components, each
containing a fraction $f_1$=$f/N$ of the total flux--density and (ii)
the magnification curves of the jet-components to be uncorrelated,
we expect the modulation-index of the combined jet-components to
decrease roughly as $\sim$$1/\sqrt{N}$. Hence, we find that
$m^{\rm obs}_\mu(N)$$\approx$$\cdot f\cdot m_\mu(f_1)/\sqrt{N}$, or
\begin{equation}
        m^{\rm obs}_\mu(f,{\cal D}) \approx \frac{C_\mu\cdot f/
        \sqrt{N}}{1+20\;\Delta\theta_{\rm b}^{-1} \; \; \sqrt{f/({\cal
        D}N)}},
\end{equation}
where $f$=$N\,f_1$$\le$1.  Hence, multiple jet-components will in
general decrease the modulation-index.  However, if the individual
jet-components are very compact -- i.e. are much smaller than the
typical separation between strong caustics --, the microlensing
variability will be dominated by single caustic crossings, creating
strong isolated peaks in the light curve. We then expect Eq.\,(34) to
break down, such that the factor $1/\sqrt{N}$ in the numerator can be
removed.

In the case of scintillation, $N$~compact jet-components
($\Delta\theta_{\rm knot}$$\la$$\theta_{\rm F}$) always vary
independently, such that the observed modulation-index roughly
decreases as $1/\sqrt{N}$. However, in the case of microlensing, the
modulation-index, caused by the same compact jet-components moving
over a magnification pattern, can be independent from $N$ or even
increase as $\sqrt{N}$. 

\subsection{Source constraints}

At this point, we give a qualitative recipe to obtain constraints on
the jet-component parameters from the observed light curves.

\begin{enumerate}

\item{First, the observed modulation-indices of both lensed images can
be used to solve for the fraction of flux density in the jet-component
($f$), as well as its Doppler factor (${\cal D}$), by comparing them
to those found from numerical simulations for different mass functions
(MFs), which fix both $C_\mu$ and $\Delta\theta_{\rm b}$). One obtains
a set of two equations (i.e. Eq.\,(33)) with two constraints ($m_\mu^{\rm
A}$ and $m_\mu^{\rm B}$) and two unknown variables (${\cal D}$ and
$f$), which in some cases can be uniquely solved for. The combinations
of MFs, which do not give a consistent solution, can be excluded,
thereby putting constraints on the allowed MFs in the line-of-sight
towards the lensed images.}

\item{Second, combining the typical observed variability time scale
between strong microlensing events and the angular separation of these
from the numerical simulations, one can obtain a value for $\beta_{\rm
app}$ (Eq.\,(27)). Combining Eqs.\,(20) and (26), one then also readily solves
for both $\psi$ and $\beta_{\rm bulk}$.}

\end{enumerate}

Thus, given (i) a mass model of the lens galaxy found from
(macroscopic) lens modeling, (ii) some plausible range of MFs near the
lensed images and (iii) the observed modulation-indices and
variability time scales in the light curves of the lensed images, one
can in principle solve for several parameters of the simple
jet/jet--component structure: ${\cal D}$, $f$, $\psi$, $\beta_{\rm
app}$, $\beta_{\rm bulk}$ and $\Delta\theta$. However, one should keep
in mind that some of the parameters might be degenerate and that we
also do not know intrinsic brightness temperature of the components.

\section{Radio Microlensing: Results}

In this section, we examine the observed variability in the
light curves of B1600+434 A and B in terms of microlensing and,
following the procedure delineated in Sect.\,4.4, derive constraints on
the properties of the jet-component, as well as on the MFs near both
lensed images. In Sect.\,7, we will use these results to determine the
microlensing modulation-index as function of frequency and compare
this to the results from independent WSRT 1.4 and 5--GHz monitoring
data of B1600+434 (Sect.\,2). An overview of the combined
microlensing/scintillation situation in B1600+434 is given in
Fig.\,\ref{situation1600}, which might act as guide to the overall
situation.

From now on, we will (i) use a source structure consisting of a static
core plus a {\sl single} relativistically moving spherically-symmetric
jet-component, (ii) assume that the core and jet parameters do not
change appreciably over the time-span of the observations, (iii)
assume that {\sl all} short-term variability in the image light curves
(Fig.\,\ref{normlc}) is dominated by microlensing and that
scatter-broadening is negligible.  The static core does not contribute
to the short-term variability, because its velocity with respect to
the magnification pattern of the lens is much smaller than that of the
jet-component (i.e. $v_{\rm core}$$\ll$$v_{\rm jet}$). At this moment,
we feel that a more detailed model is not warranted. One should,
however, keep in mind that the conclusions drawn below depend on these
assumptions\,!

\subsection{Numerical simulations}

From KBJ98, we find for B1600+434 that the local convergence and shear
are close to $\kappa$=$\gamma$=0.2 for image~A, using a singular
isothermal ellipsoidal (SIE) mass distribution for the lens galaxy.
Similarly for image~B, we find $\kappa$=$\gamma$=0.9.  Using these input
parameters, we simulate the magnification pattern of a $50\,\eta_{\rm
e}\times 50\,\eta_{\rm e}$ field for different MFs, where $\eta_{\rm
e}$ is the Einstein radius of a 1--M$_\odot$ star projected on the source
plane. We use the microlensing code developed by Wambsganss (1999).
Using the lens redshift $z_{\rm l}=0.41$, the source redshift $z_{\rm
s}$=1.59 and H$_0$=65 km s$^{-1}$ Mpc$^{-1}$ in a flat ($\Omega_{\rm
m}$=1) FRW universe, we find that $\eta_{\rm e}$=$2.1\;\mu$as. The
magnification pattern is calculated on a grid of 1000$\times$1000
pixels. Each pixel has a size of 0.107\,$\mu$as by 0.107\,$\mu$as.

We simulate 100 randomly-oriented light curves on this grid, for a
range of jet-component sizes ($\Delta \theta_{\rm
knot}$=0.125,\,0.25,..., 16.0\,$\mu$as) and MFs for the compact
objects (see Sect.\,5.2). Each simulated light curve is 54\,$\mu$as
long. For each step (5 pixels) on the light curve, we calculate the
magnification of a circular-symmetric jet-component with a constant
surface-brightness and a radius $\Delta \theta$, by convolving the
magnification pattern with its surface-brightness distribution
(e.g. Wambsganss 1999).

We also assume that the surface density near image B is dominated by
stellar objects in disk and bulge of the lens galaxy, even though the
halo does contribute to the line-of-sight surface density. If we
assume that the halo surface density near image B is equal to that
near image A ($\kappa$=0.2), the density of the caustics in the
resulting magnification pattern remains completely dominated by the
significantly higher number density of compact objects in the disk and
bulge ($\kappa$=0.7). We have tested this by modifying the
microlensing code to allow multiple mass functions. The rms
variabilities calculated from these modified magnification patterns
are the same within a few percent from those without the halo
contribution (for the range of mass functions and surface densities
that we used in this paper), which justifies this simplification.

\begin{table}[t!]
\begin{center}
\begin{tabular}{clcccc}
\hline
 MF & Mass & Slope & $\langle$M$\rangle$
	 & $C_{\mu}$ & $\Delta\theta_{\rm b}$ \\
 & (M$_\odot$) & & (M$_\odot$) & & ($\mu$as) \\
\hline
A &  \\
S1 & 0.1 & -- & -- & 0.69 & 0.94 \\
S2 & 0.2 & -- & -- & 0.70 & 1.28 \\
S3 & 0.4 & -- & -- & 0.71 & 1.87 \\
S4 & 0.6 & -- & -- & 0.69 & 2.44 \\
S5 & 0.8 & -- & -- & 0.70 & 2.48 \\
S6 & 1.0 & -- & -- & 0.71 & 2.71 \\
S7 & 1.5 & -- & -- & 0.71 & 3.07 \\
S8 & 2.5 & -- & -- & 0.69 & 4.27 \\
S9 & 5.0 & -- & -- & 0.76 & 4.50 \\
\hline
B &  \\
S1 & 0.1 & -- & -- & 1.00 & 0.90 \\
S2 & 0.2 & -- & -- & 1.01 & 1.17 \\
S3 & 0.4 & -- & -- & 1.03 & 1.41 \\
S4 & 0.6 & -- & -- & 1.05 & 1.98 \\
S5 & 0.8 & -- & -- & 1.06 & 2.35 \\
S6 & 1.0 & -- & -- & 1.07 & 2.65 \\
S7 & 1.5 & -- & -- & 1.12 & 2.95 \\
P1 & 0.01--1 & -2.35 & 0.031 & 1.01 & 1.09 \\ 
P2 & 0.01--1 & -1.85 & 0.058 & 1.05 & 1.37 \\
P3 & 0.01--1 & -2.85 & 0.021 & 0.98 & 0.71 \\
P4 & 0.1--10 & -2.35 & 0.309 & 1.09 & 2.00 \\
\hline
\end{tabular}
\end{center}
\caption{Summary of the modulation-index caused by microlensing as
function of source size for the power-law MFs (BP1--4) near image B
and the single-mass MFs (AS1--9 and BS1--7) near images A and B}
\end{table}

\subsection{The mass function of compact objects in the lens galaxy of B1600+434}

We use a range of different MFs, subdivided in two classes: 
(i) power-law MFs and (ii) single-mass MFs.

\begin{figure*}[t!]
\resizebox{8.5cm}{!}{\includegraphics{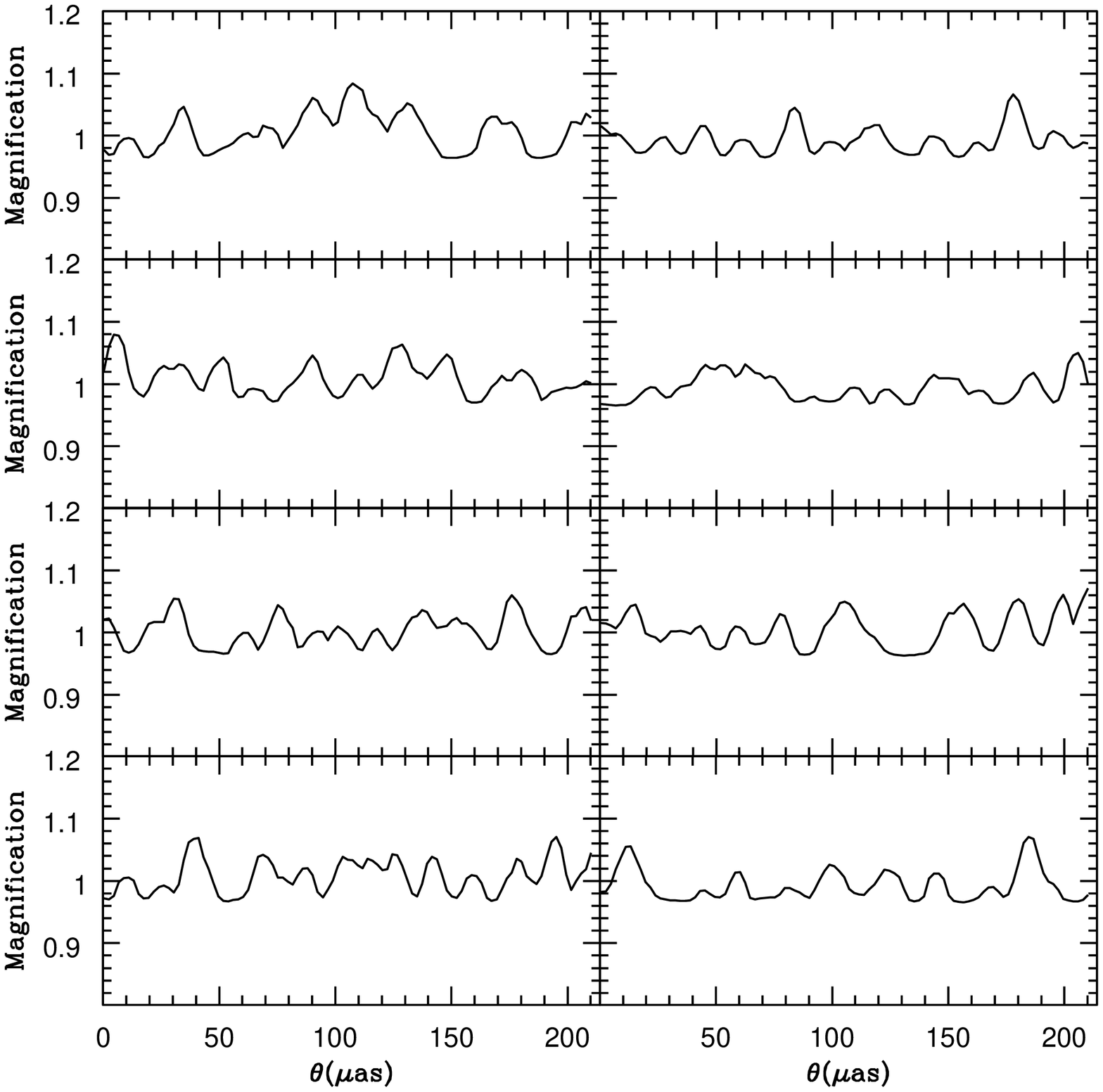}}
\resizebox{8.5cm}{!}{\includegraphics{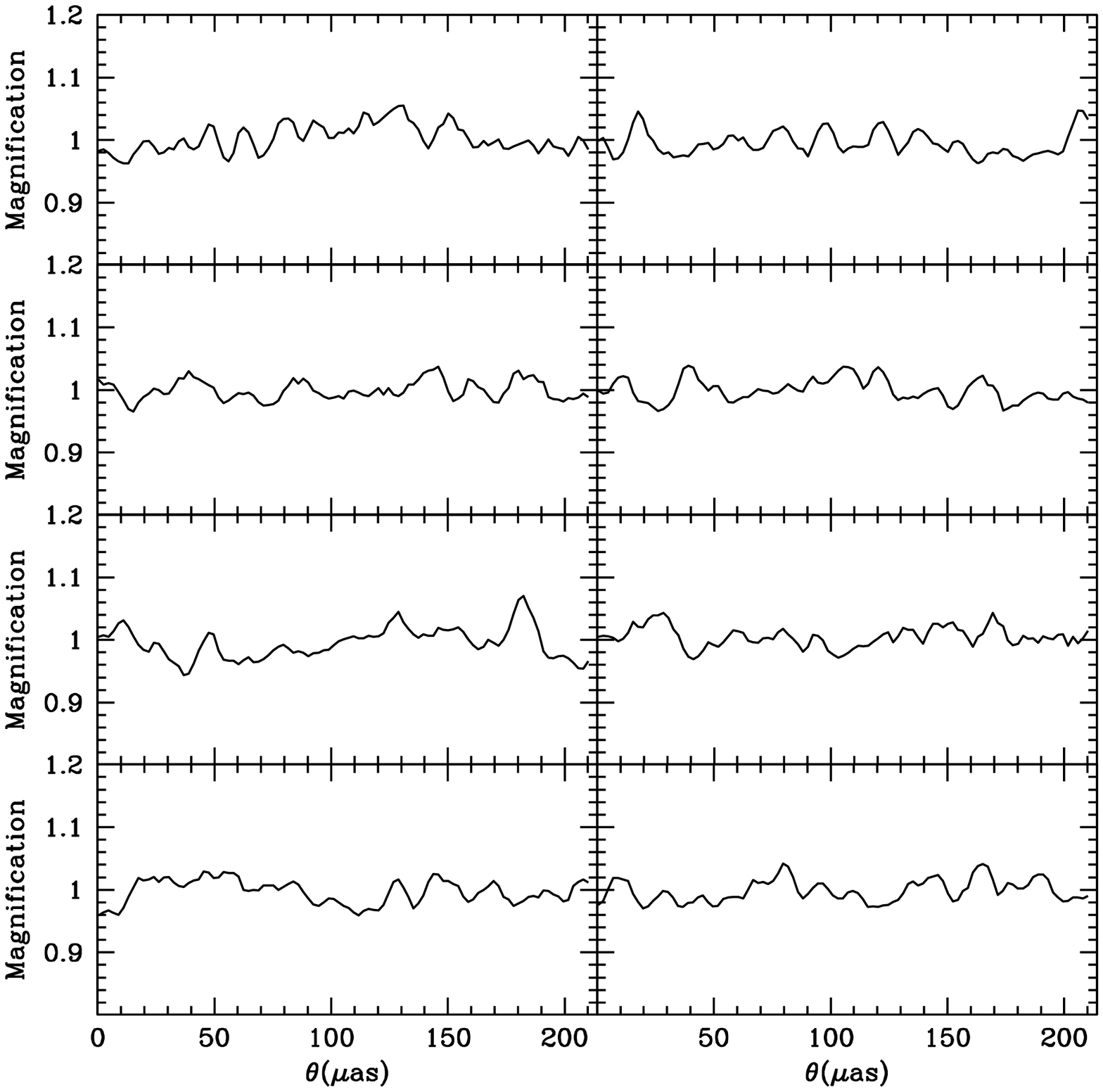}}
\vfill
\parbox[b]{\hsize}{
\caption{{\sl Left:} Eight arbitrary simulated light curves for a
single compact 5-$\mu$as source, moving over the magnification pattern
near image A.  The values $f$ and ${\cal D}$ from Table~3 were
used. The simulated curves were subsequently scaled by $f$. A
1.5--M$_\odot$ single--mass MF was used.  The light curves were
created through a convolution of the source surface--brightness
distribution with the magnification pattern. The source was moved over
a 500-pixels path (i.e. 214 $\mu$as), randomly placed on the
magnification pattern. {\sl Right:} Idem, for image~B, using a
0.2--M$_\odot$ single-mass MF. The ratio between the modulation-index
for the simulated light curves, is approximately equal to the observed
ratio of modulation-indices in the observed light curves of B1600+434
A and B (Fig.\,\ref{normlc}). The scale, in this random example, would
correspond to 35~weeks, the duration of the monitoring campaign of
B1600+434 ($\beta_{\rm app}$$\approx$18).}
\label{lcexamples}}
\end{figure*}

\subsubsection{The power-law MF}

In the solar neighborhood it appears that the stellar MF can be
represented by a single or segmented power-law of the form ${\rm
d}N(m)/{\rm d}M\propto M^{\alpha}$ (e.g. Salpeter 1955; Miller \&
Scalo 1979). Recent observations towards the Galactic bulge
(e.g. Holtzman et al. 1998) suggest a break in the MF around
0.5--0.7\,M$_\odot$, with a shallower slope at lower masses. A
lower-mass cutoff is not well constrained, although the break in the
MF suggests it might lie around a few tenths of a solar
mass. Evolution of the MF from $z$=0.4 to $z$=0.0 affects the upper
mass cutoff only and has no significant influence on microlensing,
which is dominated by the mass concentrated around the lower mass
cutoff.

\begin{enumerate}

\item{First, we simulate the magnification pattern, using a power-law
MF with $\alpha=-2.35$ (Salpeter 1955) and a mass range between
0.01--1.0\,M$_\odot$.}

\item{Second, we also investigate the slopes $-$2.85 and $-$1.85 for
image~B -- going through the bulge/disk -- assuming again a mass range
of 0.01--1\,M$_\odot$.}

\item{Third, we use an MF mass range of 0.1--10\,M$_\odot$ near image
B, with a slope of $-$2.35. This gives an average stellar mass of
about 0.3\,M$_\odot$, more in line with observations of the bulge of
our Galaxy (e.g. Holtzman et al. 1998).}

\end{enumerate}

We use the power-law MFs for image~B only, even though it is clearly
a rough approximation of the true MF. A similar power-law MF for
image~A, passing through the halo, seems unlikely, especially if the
halo consists of stellar remnants (e.g. Timmes, Woosley \& Weaver
1996).

The results from these simulations -- $m_\mu$ as function of
$\Delta\theta_{\rm knot}$ -- are fitted by Eq.\,(32).  
The values for $C_{\mu}$ and $\Delta\theta_{\rm b}$ are listed 
in Table~2 (models BP1--4).

\subsubsection{The single-mass MF}

In steep ($\alpha$$<$$-1$) MFs most of the mass is concentrated close
to the lower-mass cutoff in the MF. It therefore seems appropriate to
approximate the MF by a delta-function.  We simulated single-mass MFs
for both image~A and B, for 0.1, 0.2, 0.4, 0.6, 0.8, 1.0, 1.5
M$_\odot$ and additionally for 2.5 and 5.0--M$_\odot$ for image~A.
The results are listed in Table~2.

From these simulation, we notice several things: First, $C_\mu$ is
almost independent from the average mass of the compact objects, for a
given surface density. Second, there appears to be a strong
correlation between the average mass of the compact objects and the
turnover angle ($\Delta \theta_{\rm b}$), which means that for a given
surface density, shear and source size, the modulation-index is larger
for a higher average mass of the compact objects. The results from our
simulations are consistent with the results in Deguchi \& Watson
(1987) and Refsdal \& Stabell (1991, 1997).

\subsection{Microlensing in B1600+434?}

\begin{figure*}[t!]
\centering
\resizebox{9cm}{!}{\includegraphics{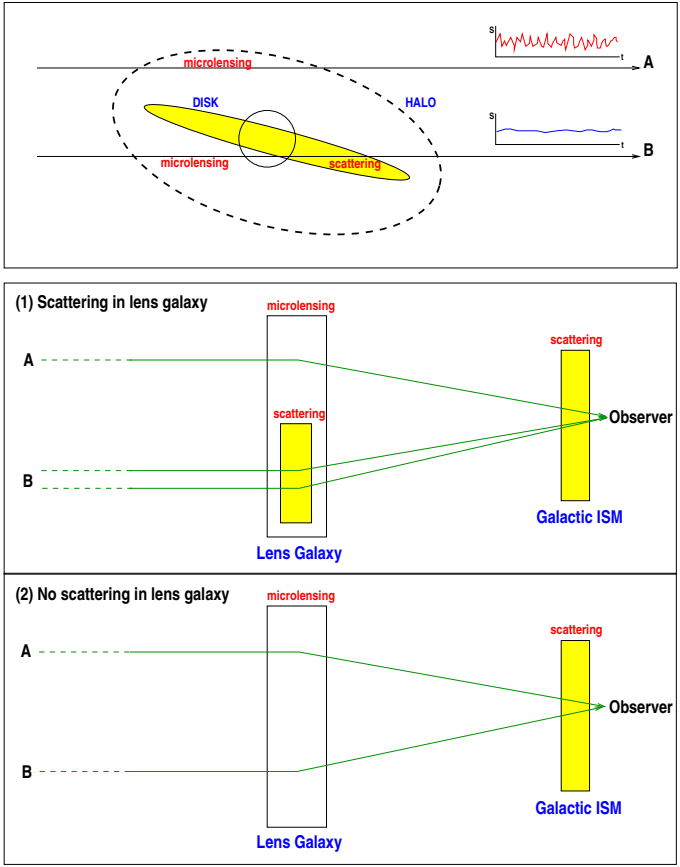}}
\hfill
\parbox[b]{8cm}{
\caption{The overall microlensing and scattering situation in
B1600+434, when seen `side-ways'. The inclination of the disk has been
slightly exaggerated ($i$=75$^\circ$). Situation 1: Light coming from
image~A passes predominantly through the halo. Light coming from image
B first passes through the bulge/disk, where the image might be
scatter-broadened by the lens--galaxy ionized ISM (Sect.\,3.2). The
image subsequently subtends a larger solid angle, suppressing
microlensing in the lens galaxy {\sl and} scintillation caused by the
Galactic ionized ISM. Situation 2: Only microlensing in the bulge/disk
and halo in the lens galaxy occurs. Subsequent scatter-broadening in
our galaxy does not suppress microlensing variability.}
\label{situation1600}}
\end{figure*}

As we can see from Table~2, the modulation-index in image~A can be
explained by a relatively small jet-component with a moderate boosting
factor. However, one might expect image~B to show similar, if not
stronger variability for a similar MF. This is not the case, however.
Before we examine this in terms of a different mass function in the
disk/bulge and halo of the lens galaxy, we first explore two
alternative explanations:

\begin{itemize}

\item{What if the dimensionless surface density near image~B is near
unity (i.e.~$\kappa$$\approx$1)? In that case, the magnification
pattern can become very dense and suppress the modulation-index
(e.g. Deguchi \& Watson 1987; Refsdal \& Stabell 1991, 1997). For
B1600+434 we used $\kappa_{\rm B}$$\approx$$\gamma_{\rm B}$
(Sect.\,5.1), in which case the density of the caustics remains
relatively small, in contrast with the case where $\kappa$$\approx$1
and $\gamma$$\approx$0 (e.g. Refsdal \& Stabell 1997). This is
supported by simulations with $\kappa=\gamma=0.999$ near image~B,
which give nearly the same results as for $\kappa=\gamma=0.9$.}

\item{In Sect.\,3.2 we showed that scatter-broadening of image B can
suppress scintillation caused by the Galactic ionized ISM. Similarly,
is can suppress variability due to microlensing.  An overview of the
situation is given in Fig.\,\ref{situation1600}.  Image~A, however, is
seen through the galaxy halo at about 4 $h^{-1}$ kpc above the disk
(KBJ98). It is therefore extremely unlikely to pass through a region
with a high scattering measure (Sect.\,3.2). In the remainder of the
paper, however, we assume that {\sl both} images A and B are not
affected by scatter-broadening (situation 2 in
Fig.\,\ref{situation1600}).}

\end{itemize}

We will now explore the only other plausible solution; a very
different MF of compact objects near images A and B.

\begin{table}[t!]
\begin{center}
\begin{tabular}{cccccc}
\hline
MF   & AS6 & AS7 & AS8 & AS9 \\
\hline
BS1    & --  & 0.09,1.3 & 0.07,2.1 & 0.05,3.4 \\
BS2    & --  & 0.10,1.2 & 0.07,2.0 & 0.05,3.2 \\
BS3    & --  & --  & 0.07,1.7 & 0.06,2.7 \\
BS4    & --  & --  & 0.08,1.5 & 0.06,2.4 \\
BS5    & --  & --  & 0.08,1.4 & 0.06,2.4 \\
BS6    & --  & --  & 0.08,1.3 & 0.06,2.0 \\ 
BS7    & --  & --  & --  & 0.07,1.4 \\
\hline
BP1    & --  & 0.10,1.2 & 0.07,2.0 & 0.05,3.2 \\
BP2    & --  & --  & 0.08,1.5 & 0.06,2.4 \\
BP3    & 0.11,1.1  & 0.09,1.6 & 0.07,2.5 & 0.05,4.0 \\
BP4    & --  & --  & 0.08,1.1 & 0.06,1.8 \\
\hline
\end{tabular}
\end{center}
\caption{All combinations of the MFs (Table~2) near image A and B
that give a consistent solution of the parameters: $f$,~${\cal D}$.
Given these two parameters, one reproduces the observed modulation
indices for both lensed images, using Eq.\,(33) and Table~3.}
\end{table}

\subsubsection{Limits on the MF and source structure}

In Table~3, we have listed all combinations between the simulated
MFs for images A and B (Table~2) that reproduce the observed
short-term modulation-indices of 2.8\% (A) and 1.6\% (B) for a
consistent set of parameters $f$ and ${\cal D}$ (Sect.\,4.4). Several
examples of simulated light curves are shown in Fig.\,\ref{lcexamples}.

\paragraph{\sf Constraints on the MFs in B1600+434:} From Table~3 one
finds that a significantly higher average mass of compact objects in
the halo is needed than in the bulge/disk to explain the modulation
indeces of both images. No consistent solutions are found for an
average mass of compact objects in the halo $<$1 M$_\odot$ for the
range of MFs that we investigated. If we furthermore put a
conservative upper limit of $\la$0.5 M$_\odot$ on the average mass of
compact objects in the bulge/disk of B1600+434 -- which lies around
the break in the Galactic bulge MF (e.g. Holtzman et al. 1998) -- only
MFs BS1--3 and BP1--4 (Table~2) remain viable MFs for the bulge/disk
of B1600+434.

\paragraph{\sf Constraints on $\Delta\theta$, $f$ and ${\cal D}$:} 
Using the MFs assumed viable above, we find from Table~3:
0.05$\la$$f$$\la$0.11 and 1.1$\la$${\cal D}$$\la$4.0.  Using Eq.\,(25)
and the values of $f$ and ${\cal D}$ listed in Table~3, the
jet-component size then lies between 2$\la$$\Delta\theta_{\rm
knot}$$\la$5\,$\mu$as.

\paragraph{\sf Constraints on $\beta_{\rm app}$:} To estimate a time
scale for strong microlensing variability, we calculate the average
power spectrum of the 100 light curves for each MF, for the source
size of 2 and 4 $\mu$as. The power-spectrum is typically relatively
flat at low frequencies, smearing out the long-period modes in the
light--curves. At higher frequencies the power drops rapidly. We
therefore expect the strongest Fourier modes in the light curves to
lie around the turn-over frequency, where the power drops to about
50\%. Consequently, we define the typical angular scale of variability 
($\theta_{\rm t}$) to correspond with the half-power frequency in the
power-spectrum. In Table~4, we listed the results for those MFs that
give a consistent solution (Table~3).
 
If we now take a separation of $\sim$2 weeks as indicative for the
separation of strong modulations in light curve of image~A (see days
80--140 in Figs \ref{normlc}--\ref{normdiff}), we find an angular
velocity of the jet-component in the source plane between 3 and 
9\,$\mu$as/week. Using Eq.\,(30), we then derive 9$\la$$\beta_{\rm
app}$$\la$26.  This range strongly dependents on the local structure
of the magnification pattern, which can differ strongly from place to
place (e.g. Wambsganss 1990).

\begin{figure}[t!]
\center 
\resizebox{\hsize}{!}{\includegraphics{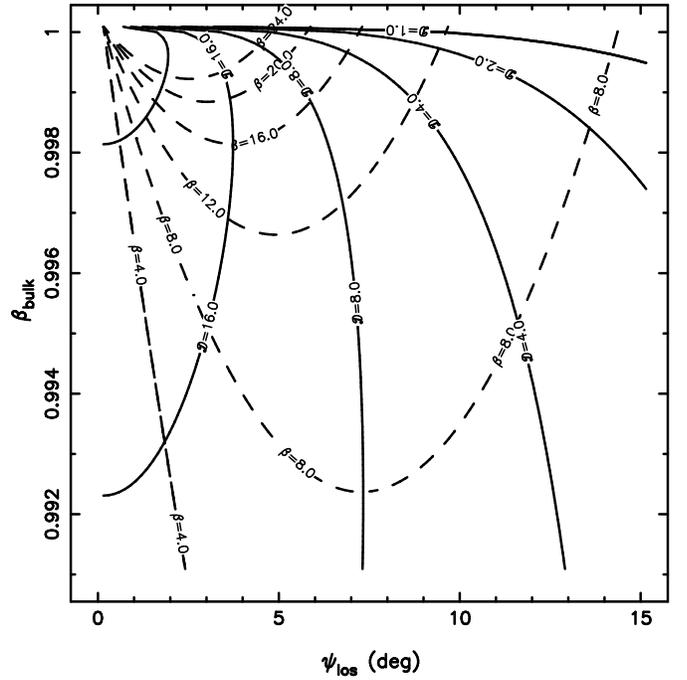}}
\hfill
\parbox[b]{\hsize}{
\caption{The bulk velocity ($\beta_{\rm bulk}$) and angle with respect to the
the line-of-sight to the observer ($\psi$) of a relativistic jet-component,
as function of its Doppler boosting (${\cal D}$) and apparent 
velocity ($\beta_{\rm app}$), calculated using Eqs.\,(20) and (26).}
\label{psibeta}}
\end{figure}

\paragraph{\sf Constraints on $\beta_{\rm bulk}$ and $\psi$:} 
Using the constraints of the Doppler factor (${\cal D}$) and angular
velocity of the jet-component ($\beta_{\rm app}$), one also obtains
constraints on the angle between the jet--component direction with
respect to the line-of-sight to the observer and the bulk velocity of the
jet-component ($\beta_{\rm bulk}$). From Fig.\,\ref{psibeta}, we
subsequently find: $4^{\circ}$$\la$$\psi$$\la$$13^{\circ}$ and
$\beta_{\rm bulk}$$\ga$$0.995$, for the allowed ranges of these parameters.

Thus, several combinations of MFs for images A and B (Table~3) give
solutions that reproduce the observed modulation-indices of both
images for a consistent, although not unique, set of jet-component
parameters. The derived constraints on the jet-component, however,
do agree with observations of confirmed superluminal sources (e.g. 
Vermeulen \& Cohen 1994).

\begin{table}[t!]
\begin{center}
\begin{tabular}{ccc}
\hline
MF   &  $\Delta\theta_{\rm knot}$=2 $\mu$as & 4 $\mu$as \\
\hline
AS6    & 8  & 14 \\
AS7    & 8  & 13 \\
AS8    & 10 & 14 \\
AS9    & 13 & 14 \\
\hline
BS1    & 7 & 13 \\
BS2    & 7 & 12 \\
BS3    & 7 & 12 \\
BS4    & 8 & 14 \\
BS5    & 10 & 14 \\
BS6    & 8 & 15 \\
BS7    & 8 & 18 \\
\hline
BP1 & 7 & 14 \\
BP2 & 6 & 12 \\
BP3 & 7 & 16 \\
BP4 & 7 & 12 \\
\hline
\end{tabular}
\end{center}
\caption{The typical angular scale ($\theta_{\rm t}$) in $\mu$as
between strong microlensing events in the simulated light curves, as
defined through the power spectrum (see text). Listed are the values
for two source sizes, 2 and 4\,$\mu$as, approximately corresponding to
the range of jet-component sizes that reproduce the observed
modulation-index in the light curves (see text).}
\end{table}

\subsubsection{Microlensing by compact halo objects}

It appears we have found a lower limit of $\sim$1 M$_\odot$ on the
mass of compact objects in the halo around the lens galaxy, under the
assumptions that all variability we see is due to microlensing, the
jet is dominated by a single component and there is no
scatter-broadening.  Let us now explore the implications of this in
more detail, first mentioning several potential problems.

% Checked upto here...

\begin{itemize}

\item{Could microlensing be due to a globular cluster (GC) in the halo
of the lens galaxy? It is easy to show that the probability of seeing a lensed image
through a GC surface density $>\kappa_{\rm GC}$ is
\begin{equation}
	P(>\kappa_{\rm GC})\approx\frac{N}{4\cdot\kappa^2_{\rm GC}}
	\left(\frac{\sigma_{\rm GC}}{\sigma_{\rm LG}}\right)^4.
\end{equation}
We take a population of $N$$\sim$150 GCs inside the Einstein radius of
the lens galaxy (LG), with a velocity dispersion of $\sigma_{\rm
GC}$=7\,km~s$^{-1}$. These values are typical of those found for our
galaxy (e.g. Binney \& Merrifield 1999).  Using $\sigma_{\rm
LG}$$\approx$200 km s$^{-1}$ (KBJ98), $\kappa_{\rm GC}$=$\kappa_{\rm
A}$=0.2, we then find a probability $P(>\kappa_{\rm
GC})$$\approx$$1.7\cdot 10^{-3}$ that the microlensing optical depth
($\tau_{\rm ml}$$\approx$$\kappa_{\rm ml}$) of the GC exceeds that of the
dark matter halo. The probability that $\kappa_{\rm GC}$ exceeds
$\kappa_{\rm B}$=0.9, thereby causing similar or larger microlensing
variability, is $7\cdot10^{-5}$. Hence, it is very unlikely that a GC
in the line-of-sight to lensed image~A could enhance the microlensing
optical depth significantly.}

\item{What is the influence of binary systems on the mass limit of a
compact object in the dark matter halo? We know that a large fraction of stars
in the bulge is locked up in binary systems (e.g Holtzman et
al. 1998). In the case of high microlensing optical depths, one can
consider a binary as a single microlensing object, with a mass equal
to the sum of the individual masses.

From Table~2 we see that a higher mass of compact objects gives a
higher microlensing modulation-index for a given source size. Hence,
if we know the typical stellar mass of objects in the bulge/disk and
assume they are all single objects, not in a binary system, we
underestimate its modulation-index. One has to take this effect into
account.}

\end{itemize}

Having dealt with these possible complications, let us explore the
lower mass limit of the compact objects in more detail.  If we assume that (i)
all compact object in the bulge/disk are not in binary systems and
(ii) all compact objects are in binary systems and (iii) use the lowest average
mass of objects in the bulge/disk (Tables 2--3), we find a very
conservative lower limit of 0.5 M$_\odot$ on the mass of individual
compact objects in the halo.  In the more realistic case where most of the
stars in the bulge/disk are in binaries and most compact objects are probably
not, a lower limit of 1.5 M$_\odot$ is found, assuming that the
average bulge/disk stellar mass in the halo is $\sim$0.1 M$_\odot$. 
If the bulge/disk stars have average masses somewhere between 0.1 and 
0.3 M$_\odot$ and are foremost in binaries, the lower limit increases 
to $\sim$2.5 M$_\odot$.

As in the case of scintillation, scatter-broadening of image B also
suppresses microlensing (Sect.\,5.3). If this happens, one would
underestimate the true microlensing modulation-index of image B. This
would give one more freedom to decrease $f$ and/or increase ${\cal D}$
for the microlensed jet-component, thereby changing the required
average mass of compact objects in the halo. It would, however, {\sl never}
eliminate the need for them.

\begin{table}[t!]
\begin{center}
\begin{tabular}{llc}
\hline
Fraction flux             & $0.05\la f\la 0.11$ \\
Doppler factor            & $1.1\la {\cal D}\la 4.0$ \\
Size in source plane      & $2\la \Delta\theta_{\rm knot}\la 5$ & ($\mu$as)\\
Apparent velocity         & $9\la \beta_{\rm app}\la 26$ \\
LOS angle of jet          & $4\la \psi_{los}\la 13$     & (${^\circ}$)\\
Bulk velocity             & $\beta_{\rm bulk}$$\ga$0.995\\            
\hline
\end{tabular}
\end{center}
\caption{Summary of constraints on the jet-component parameters,
derived within the context of the microlensing hypothesis. We assume
T$_{12}^{\rm b,ic}$=0.5 and a flat FRW universe with $\Omega_{\rm
m}$=1 and H$_0$=65 km s$^{-1}$ Mpc$^{-1}$.}
\end{table}

\begin{figure}[t!]
\center 
\resizebox{\hsize}{!}{\includegraphics{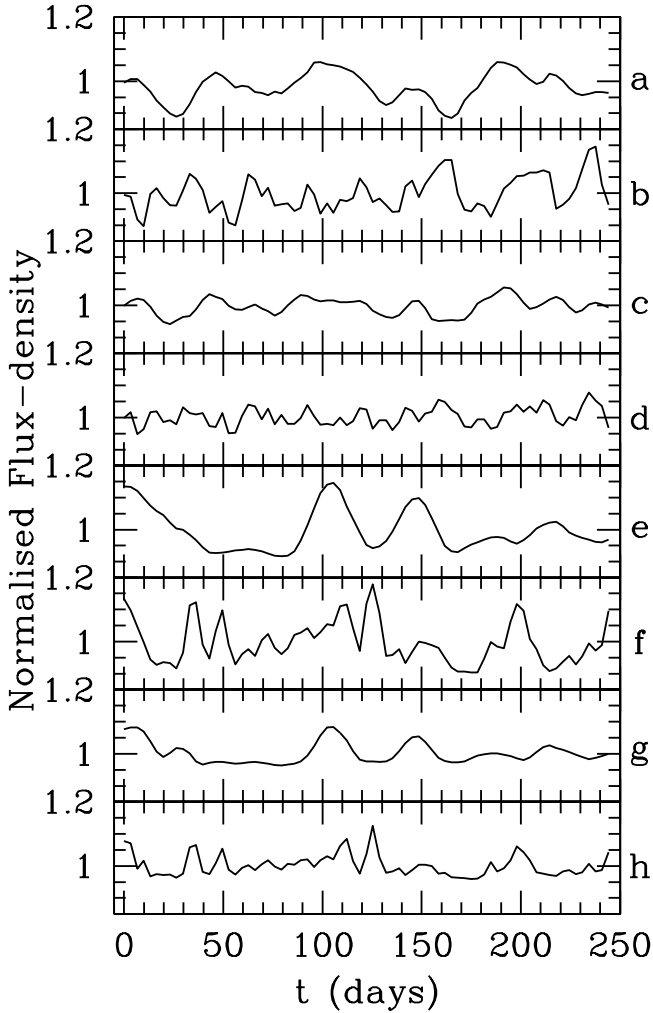}}
\vfill
\parbox[b]{\hsize}{
\caption{Eight simulated microlensing light curves of 3C120.
The parameters for each light curve are listed in Table~6.}
\label{lc3c120}}
\end{figure}

\section{Microlensing of a realistic jet structure}

So far, we have only investigated a very simple model of the source
structure, consisting of a core plus a single jet-component.  To see how a real
source behaves, when microlensed by similar MFs as in the lens galaxy
of B1600+434, we simulated light curves of a more complex jet structure.

\subsection{The jet structure of 3C120}

We used the inner jet-structure seen in 3C120 (G\'omez et al. 1998;
G\'omez, Marscher \& Alberdi 1999) as a `template', because it is one
of the best-studied nearby jet structures. The inner jet consists of
a core plus at least 15 distinguishable jet-components (G\'omez et
al. 1999), which move superluminally, with velocities typically around
4.5$c$ (H$_0$=65 km s$^{-1}$ Mpc$^{-1}$; G\'omez et al. 1998). The jet was observed at 22
GHz. The total flux density of 3C120 at this frequency is 5.7$\pm$0.9
Jy (O'Dell et al. 1978), whereas the fitted inner-jet-components
contain about 2.5 Jy (G\'omez et al. 1999). 
We therefore assume that the inner jet contains about 40\%
of the total flux density of 3C120.

Note that 22\,GHz corresponds to approximately X--band (8.5\,GHz)
observations, when placed at the source redshift of B1600+434
($z$=1.59). We thus scale the size of the jet structure down by the ratio
of angular diameters distances for 3C120 and the lensed source of
B1600+434 and its flux density by the ratio of luminosity distances
squared.

At a redshift of $z$=1.59, 3C120 would be a 1.8\,mJy source. To obtain a
source of about 25\,mJy total , as observed for B1600+434, we scale each jet
component's size by a factor $\alpha_{\rm r}$=3.7 in radius
and assume that about 60\% of its flux density is contained in an
extended structure, which is not sensitive to microlensing. We assume
their surface brightness temperature to be conserved and associate the
radius of the source with half the FWHM of the component size
determined by G\'omez et al. (1999). The resulting jet structure has a
flux density of about 25\,mJy at 8.5\,GHz and an angular size of the
inner jet less than 1 mas. If we only use the inner jet and not
assume the additional 60\% of extended emission, one should scale the
inner-jet-components by a factor of $\alpha_{\rm r}$=5.7 to obtain a
total flux density of about 25\,mJy.

\subsection{Microlensed light curves of 3C120}

The jet structure is randomly placed on the magnification pattern.  We
recalculate the jet structure and the resulting normalized light
curves at epochs separated by 3.3 days, which is the average sampling
of the light curves of B1600+434. For image~A, we use the MF AS7,
which corresponds to a halo filled with stellar remnants, such as
black holes and neutron stars. For image~B, we use the MF BS2.

We calculate light curves with a total time span of 35 weeks,
corresponding to the length of the observed VLA light curves of
B1600+434. We subsequently scale the light curves by a factor 0.4
($\alpha_{\rm r}$=3.7 for 40\% of the flux density from the inner
jet) or 1.0 ($\alpha_{\rm r}$=5.7 for 100\% of the flux density from
the inner-jet). We repeat these simulations using an apparent
velocity three times larger ($\alpha_{\rm v}$=3).

In Fig.\,\ref{lc3c120}, one simulated light curve is shown for the
images A and B, for each size scale factor ($\alpha_{\rm r}$) and
velocity scale factor ($\alpha_{\rm v}$). The modulation-index and
estimated variability time scale between strong microlensing events
are listed in Table~6.

\begin{table}[t!]
\caption{Scaling parameters for 3C120 and results from eight
arbitrary simulated light curves. LCs $a-d$ show light curves
using the MF B2, whereas LCs $e-h$ show light curves using the MF 
A7.}
\begin{center}
\begin{tabular}{lcccc}
\hline
LC  &  $\alpha_{\rm r}$ & $\alpha_{\rm v}$ & rms (\%) & $t_{\rm typ}$ (d) \\
\hline
a   & 5.7 & 1 & 4.5 & $\sim$50\\
b   & 5.7 & 3 & 5.4 & $\sim$15\\
c   & 3.7 & 1 & 2.6 & $\sim$35\\
d   & 3.7 & 3 & 2.9 & $\sim$10\\
\hline
e   & 5.7 & 1 & 6.4 & $\sim$50\\
f   & 5.7 & 3 & 6.7 & $\sim$20\\
g   & 3.7 & 1 & 3.3 & $\sim$40\\
h   & 3.7 & 3 & 3.4 & $\sim$15\\
\hline
\end{tabular}
\end{center}
\end{table}

\subsection{A comparison with B1600+434}

Not only does the modulation-index correspond well with that seen for
B1600+434--A and B, also the time scale of variability is in the order
of several weeks to months (depending on the choice of $\alpha_{\rm
v}$; Table~6). The similarity between some of the simulated light
curves and those observed for B1600+434 is remarkable, knowing that we
have not resorted to extreme assumptions.

We therefore regard these simulations as `proof of principle', showing
that microlensing of multiply-imaged compact flat-spectrum radio
sources, of which more and more are being discovered -- for example in
the CLASS/JVAS survey (e.g Browne et al.  1997; Myers et al. 2000) --
can be a very common occurrence, enabling us to study both the
structure of these high--$z$ radio sources, as well as the MF of
compact objects in the intermediate--$z$ lens galaxies.

\section{Microlensing versus Scintillation}

We have seen many arguments in Sections 3--5, for and/or against
scintillation and microlensing as the cause of variability in compact
flat-spectrum radio sources. Both can in principle explain the
observed modulation-index and variability time scales in the VLA
8.5--GHz light curves in the individual lensed images of B1600+434,
although it remains very difficult to explain the longer ($\gg$1 day)
variations in the light curves or the difference in modulation-index
between the two lensed images in terms of scattering. In the case of
microlensing, one would expect to see scintillation at some level as
well, possibly complicating a straightforward analysis.  How can we
then separate these mechanisms as the dominant cause of variability?

Scintillation and microlensing have different dependencies on
frequency. Although microlensing is achromatic, the frequency
dependence of the source structure predicts a clear dependence of the
microlensing variability as function of frequency. For flat-spectrum
synchrotron self-absorbed source sources, the source size is inverse
proportional to frequency (Eq.\,(24)). Thus the modulation-index
decreases with decreasing frequency (Eq.\,(32)). In the case of weak and
strong refractive scattering (i.e. flickering), however, the
modulation-index usually increases with decreasing frequency
(e.g. BNR86; Narayan 1992; Rickett et al. 1995). {\sl The key to
testing whether the observed short-term variability in
gravitationally lensed flat-spectrum radio sources is partly or fully
dominated by microlensing or scintillation is therefore their strong
opposite dependence on frequency.}

\begin{figure}[t!]
\center 
\resizebox{\hsize}{!}{\includegraphics{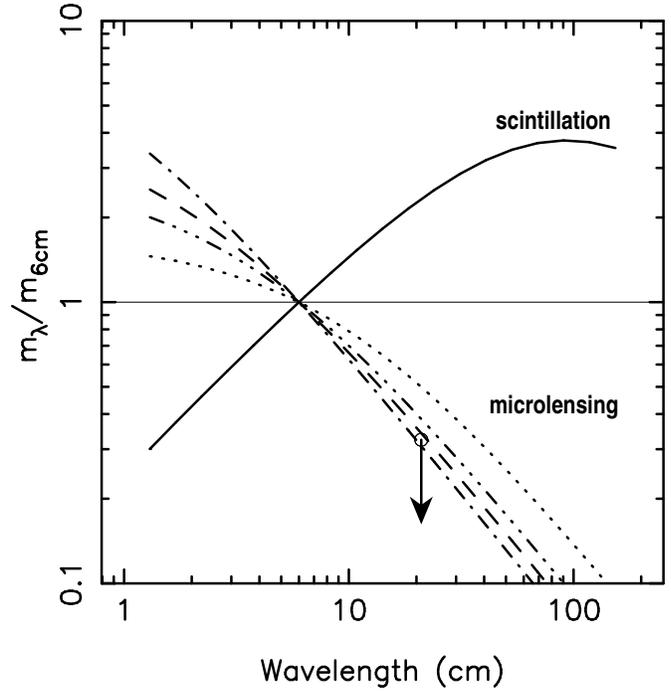}}
\hfill
\parbox[b]{\hsize}{
\caption{Dependence of the modulation-index from scintillation and
microlensing on wavelength. The solid line shows the modulation-index
from scintillation. The dashed and dot-dashed lines show the
modulation-indices from microlensing, using $\Delta\theta_{\rm
knot}$=2 and 5 $\mu$as at 8.5\,GHz, respectively, assuming
$\Delta\theta_{\rm b}$=0.9 $\mu$as. The dotted and dash-dot-dotted
lines indicates the same for $\Delta\theta_{\rm b}$=4.5 $\mu$as. All
curves are normalized to the $m_{\rm part}$=3.7\% modulation at 6.0 cm
observed with the WSRT in 1999 (Table~1). The open circle indicates 
the upper limit on the ratio $m_{21}/m_{6}$ (Table~1).}
\label{predict}}
\end{figure}

In Fig.\,\ref{predict}, we have plotted the dependence of the modulation
index in the case of weak and strong refractive scintillation versus
that of microlensing. We assume that the source or jet-component size
scales as $\nu^{-1}$ and that scatter-broadening is negligible. All
curves are normalized at $m_{\rm part}$=3.7\% at 6 cm, as measured
with the WSRT in 1999 (Table~1). We determine the modulation-index
from scintillation following Rickett et al. (1995) and that from
microlensing using Eq.(32). In the case of microlensing, we use the
maximum range of the turn-over scale $\Delta\theta_{\rm
b}$=0.9--4.5~$\mu$as and the jet-component size at 8.5\,GHz
$\Delta\theta_{\rm knot}$=2--5~$\mu$as, found from Tables 2 and 5. We
furthermore assume that the short-term variability is dominated by
image~A, as observed in the VLA 8.5--GHz light curves. The resulting
curves show a clear opposite trend as function of wavelength and
therefore act as a strong discriminator between microlensing and
scintillation. The constraints on the microlensing curves were 
determined from the VLA 8.5--GHz light curves only and therefore
independent from the WSRT 1.4 and 5--GHz modulation-indices. 

From the WSRT modulation-indices ($m_{\rm part}$) at 1.4 and 5\,GHz
(Table~1), one finds $m_{21}/m_6$$\approx$0.31, as indicated by the
open circle in Fig.\,\ref{predict}. Although this result is based on
two frequencies only, it clearly agrees much better with the
predictions from microlensing and not that from scintillation! The
latter would require a $\sim$8 times larger value for $m_{21}$ (i.e
about 9\%). We do not plot the VLA 8.5--GHz modulation-index, because
it was determined from a different epoch. The modulation-index from
microlensing and scintillation might change as function of time,
whereas the ratio of modulation-indices, measured simultaneously, is
less likely to change.

Thus, if all short-term variability seen in the light curves of
B1600+434 is external, it follows the predictions from
microlensing. Moreover, if the short-term variability seen in the
WSRT 1.4 and 5--GHz light curves is intrinsic, this would be very hard
to reconcile with the fact that in 1998 almost all of the VLA 8.5--GHz
short-term variability was shown (see
Sect.\,2) to be external. The most logical conclusion from all this is
that the short-term variability at 1.4, 5 and 8.5\,GHz is dominated
by microlensing. This explains both the modulation-indices as function
of frequency at 1.4 and 5\,GHz and the longer variability time scale
in the VLA 8.5--GHz light curves. In the case of scintillation, one
would require either different sizes of the lensed images or a very
different ionized ISM towards the images and also a different
time scale and frequency-dependence of scattering from that expected
from a Kolmogorov spectrum of inhomogeneities of the ionized ISM. All
evidence thus far is therefore {\sl only} consistent with microlensing
as the dominant cause of the observed short-term variability.

\section{Summary \& Conclusions}

We have shown {\sl unambiguous} evidence of external variability in
the CLASS gravitational lens B1600+434. The difference between the
8.5--GHz VLA light curves of the two lensed images shows external
variability at the 14.6--$\sigma$ confidence level. The modulation
indices of the short-term variability are 2.8\% for image~A and 1.6\%
for image~B. The difference light curve has an rms scatter of 2.8\%,
indicating that the short-term variability in both light curves is
mostly of external origin (Sect.\,2).

We have investigated two plausible sources of this external
variability: (i) scattering by the ionized component of the Galactic
interstellar medium (ISM) and (ii) microlensing by massive compact
objects in the bulge/disk and halo of the lens galaxy.

Based on the `standard' theory of scintillation (e.g. Narayan 1992;
Rickett et al. 1995)
there should be a considerable increase in the modulation-index with
wavelength (Sections 3 and 7). From simultaneous WSRT 1.4 and 5--GHz
observations we find, however, that $m_{21}$=1.2\% and $m_{6}$=3.7\%
(Table~1), which is a considerable decrease. Scintillation theory
predicts $m_{21}$=9.0\% for $m_{6}$=3.7\% (Sect.\,7). If the 1.4 and
5--GHz short-term variability is intrinsic, it is hard to reconcile
with the fact that in 1998 the VLA 8.5--GHz light curves were
dominated by external variability during the full eight months of
monitoring (Sect.\,2), although it can not be fully excluded yet.
Moreover, from microlensing simulations, we expect that
$m_{21}$=1.2--2.4\% if $m_{6}=3.7\%$ (Fig.\,\ref{predict}), based on
constraints on the source structure and mass function of compact
objects found from the VLA 8.5--GHz light curves (Sections 4, 5 and
7).  This range agrees remarkably well with the observed modulation
index $m_{\rm part}$=1.2\% at 21 cm.

Supplementary to this argument, the difference in modulation-index
between the lensed images would, in the case of scintillation, argue
for either a very different Galactic ionized ISM (SM$_{\rm
A}/$SM$_{\rm B}\approx$3.1; Section 3.1--2) towards the lensed images
or a different image size ($\Delta\theta_{\rm B}/\Delta\theta_{\rm
A}$$\approx$1.75; Sect.\,3.2), although the latter might result from
scatter-broadening. Furthermore, the longer variability time scales
at 8.5\,GHz ($\gg$1 day; Figs \ref{normlc}--\ref{normdiff}) are also
difficult to explain in terms of scintillation, as well as the absence
of variability with short time scales in several 12\,h WSRT
observations at 5\,GHz (Koopmans et al. in prep.).

However, the strongest argument against scintillation remains the
dominant presence of short-term external variability at 8.5\,GHz in
1998, combined with the fact that in 1999 significant short-term
variability is seen at 5\,GHz, but almost none at 1.4\,GHz.

Under the microlensing hypothesis, we find a consistent, although not
unique set of jet-component parameters. A core plus a
single-jet-component with a size of 2--5 $\mu$as, containing 5--11\%
of the flux density and moving superluminally with 9$\la$$\beta_{\rm
app}$$\la$26, can explain the modulation-index and variability time
scale in both lensed images (Sections 4--5). For image~A we find a
significantly higher average mass of compact objects ($\ga$0.5
M$_\odot$), compared with those near image~B.  A much lower mass of
compact object would result in a finer magnification pattern and thus
in less variability. If image~B is scatter-broadened, its
microlensing modulation-index is reduced, which might change the
lower-limit on the compact object mass.

If one, based on the evidence gathered thus far, accepts that the 1.4,
5 and 8.5--GHz short-term variability in B1600+434--A and B is
dominated by microlensing, the profound consequence is that the
dark-matter halo at $\sim$6 kpc above the plane of the disk-galaxy
lens in B1600+434 is partly filled with massive compact objects. New
WSRT, VLA and VLBI multi-frequency data is being obtained at the
moment, which combined with a more comprehensive statistical analysis 
should provide us with refined constraints on the mass function of 
compact objects and the source structure (Koopmans et al. in prep.).

\begin{acknowledgements}

The authors would especially like to thank Joachim Wambsganss for
making available and giving support in using his microlensing code.
They thank Frank Briggs for providing useful supermongo code, and
Joachim Wambsganss, Jane Dennett-Thorpe, Penny Sackett, Jean-Pierre
Macquart and Roger Blandford for useful discussions. They also thank
the referee, Andreas Quirrenbach, for pointing out several important
issues. LVEK and AGdeB acknowledge the support from an NWO program
subsidy (grant number 781-76-101). This research was supported in part
by the European Commission, TMR Programme, Research Network Contract
ERBFMRXCT96-0034 `CERES'. The National Radio Astronomy Observatory is
a facility of the National Science Foundation operated under
cooperative agreement by Associated Universities, Inc.  The National
Radio Astronomy Observatory is a facility of the National Science
Foundation operated under cooperative agreement by Associated
Universities, Inc. The Westerbork Synthesis Radio Telescope (WSRT) is
operated by the Netherlands Foundation for Research in Astronomy
(ASTRON) with the financial support from the Netherlands Organization
for Scientific Research (NWO).

\end{acknowledgements}

\end{document}